\documentclass[english,aps,pra,twocolumn,superscriptaddress]{revtex4-1}
\usepackage[latin9]{inputenc}
\usepackage{amsmath}
\usepackage{amssymb}
\usepackage{graphicx}

\makeatletter
\usepackage[colorlinks,citecolor=blue,linkcolor=blue]{hyperref}
\usepackage{amsmath}
\allowdisplaybreaks[2]

\makeatother

\usepackage{babel}
\begin{document}
\title{Enhanced Quantum Parameter Estimation via Dynamical Modulation}
\author{Guohui Dong}
\email{dongguohui@sicnu.edu.cn}

\affiliation{College of Physics and Electronic Engineering, Sichuan Normal University,
Chengdu 610068, China}
\author{Yao Yao}
\email{yaoyao_mtrc@caep.cn}

\affiliation{Microsystem and Terahertz Research Center, China Academy of Engineering
Physics, Chengdu 610200, China}
\begin{abstract}
In quantum theory, the inescapable interaction between a system and
its surroundings would lead to a loss of coherence and leakage of
information into the environment. An effective approach to retain
the quantum characteristics of the system is to decouple it from the
environment. However, things get complicated in quantum parameter
estimation tasks, especially when the unknown parameter originates
from the environment. Here, we propose a general dynamical modulation
scheme that enables the enhancement of estimation precision irrespective
of the origin of the parameters to be estimated. Notably, beyond the
capability of dynamically decoupling the system from the environment,
our proposal can select the effective frequency of the system and
in turn manipulate the dynamics of the system, such as the steady-state
probability and the decay rate, which are sensitive to the bath parameter.
Therefore, the quantum Fisher information is dramatically increased,
leading to superior estimation precision. Specifically, we elucidate
the effectiveness of our proposal with prototypical examples of the
Ramsey spectroscopy and quantum thermometer. Our findings offer an
alternative route to enhance the precision of quantum estimation processes
and may have potential applications in diverse branches of quantum
metrology.
\end{abstract}
\maketitle

\section{Introduction}

In reality, any open quantum system unavoidably interacts with its
environment (also deemed as the reservoir or heat bath). With this
environment-induced noise, the quantum system would gradually lose
its coherence (the decoherence process) \citep{2002open_system},
which prevents the system from quantum tasks such as quantum computation,
quantum communication, and quantum teleportation \citep{1995quantum_computation,2010Nielsen,2015quantum_communication}.
Therefore, how to effectively suppress the decoherence caused by the
environment has long been a central challenging topic in open quantum
systems. From the perspective of quantum information theory, two lines
of thought have been developed to mitigate the effect of ambient noise.
One is the passive approach that embeds the quantum system into a
decoherence-free subspace. In such subspace, the system is immune
to noise due to symmetry \citep{1997Zanardi,1997Duan,1998decoherence_free_subspace,2000decoherence-free-subspace}.
The other is the active approach. For instance, the quantum error-correcting
codes protect the information in quantum states from noise by using
redundant qubits \citep{1995Shor,1996Calderbank,2010Nielsen}. As
another representative example, the dynamical modulation (DM) technique
exploits an external electromagnetic field (continuous wave or sequence
of pulses) to control the dynamics of the system. When the frequency
of the field is larger than the inverse of the reservoir's memory
time, the effective coupling strength between the system and the environment
will be manipulated. Hence, for a judicious choice of modulation parameters,
the undesired system-environment interaction can be washed out dynamically,
suppressing the decoherence of the system \citep{1998SethLloyd,1999Agarwal,2007Fanchini,2013tanqingshou,2022Cai_dynamical_coupling_machine_learing}.

Recently, quantum metrology, which aims at enhancing measurement precision
with quantum characteristics such as quantum entanglement and squeezing,
has attracted much attention \citep{2006quantummetrology,2011quantummetrology,2017quantum_sensing}.
In a quantum parameter estimation scenario, the parameter of interest
is first encoded into the quantum state of the system (the parametrization
process). Then a series of measurements are performed to extract the
information stored in this state. Roughly speaking, according to the
origin of the parameters in the parametrization process, there are
two types of estimation: (i) the parameter independent of the environment
(estimation type I hereinafter), for example, the atomic spectroscopy
\citep{1994atomic_spectroscopy,1997atomic_spectroscopy,2017atomic_spectroscopy}
and interferometry \citep{2004interferometry,2012interferometry};
(ii) the parameter from the environment (estimation type II hereinafter),
such as the heat-bath temperature \citep{2010thermometry,2015thermometry,2018Campbell_thermometer,2018thermometry,2019thermometry,2022thermometry}
and the spectral density of the reservoir \citep{2017bath_sensing,2018bath_sensing,2019bath_sensing,2020bath_sensing,2021bath_sensing}.

Due to the different roles that the environment plays in two estimation
types, the proposals enhancing the estimation precision should be
designed appropriately. For the estimation type I, the environment
only induces the information leakage of the system and thus reduces
the estimation precision \citep{2011Escher,2011Ma,2017quantum_sensing}.
Therefore, the passive and active decoherence-suppression protocols
are both effective precision-promotion approaches \citep{2006Roos,2022Hamann,2015phaseestimation}.
However, things get complicated when one deals with the estimation
type II where the system-environment interaction is essential in the
parametrization process. Obviously, the passive scheme which physically
decouples the system from the environment can not achieve precision
improvement. Hence, a natural question arises in our mind: Whether
the active approach, i.e., the dynamics modulation, is valid for the
estimation type II?

In this paper, we propose a general dynamical modulation scheme that
enables the enhancement of estimation precision irrespective of the
origin of the parameters. More specifically, with the aid of a time-varying
control field, we explore the parameter estimation processes in two
estimation types with characteristic examples of the Ramsey spectroscopy
and quantum thermometer. Beyond the capability of dynamically decoupling
the system from the environment, our proposal can select the effective
frequency of the system and in turn manipulate the dynamics of the
system, such as the steady-state probability and the decay rate, which
are sensitive to the bath parameter. By demonstrating the dramatically
increased Fisher information, we verify the effectiveness of the dynamical
modulation as one alternative route for the precision enhancement.

The remainder of this paper is organized as follows. In Sec. \ref{sec: Parameter-estimation-and-Fisher-information},
we briefly summarize the basic concepts of estimation precision and
the fundamentals of Fisher information. In Sec. \ref{sec:DDT-dynamics},
we first explore the DM in an open quantum system and then illustrate
two features of the DM, i.e., the shrinkage of the system-environment
coupling strength and the effective frequency selection of the system.
Taking the Ramsey spectroscopy and quantum thermometer as illustrative
examples, we elucidate the underlying physics of the DM-enhanced estimation
precision in two types respectively in Sec \ref{sec:enhanced-parameter-estimatio}.
Section \ref{sec:Discussions-and-conclusions} provides a summary
and discussion of our results. Details of the calculations are given
in Appendixes \ref{sec:integro-differential equation}-\ref{sec:appen-bosonic-bath}.

\section{Quantum Parameter estimation and Fisher information\label{sec: Parameter-estimation-and-Fisher-information}}

As a preliminary introduction to quantum parameter estimation, we
present a brief summary of the basic concepts of the estimation precision
and the fundamentals of the Fisher information.

In the classical estimation theory, an observable X carries the information
of an unknown real parameter $\varphi$. $\{p_{i}(\varphi)\}$ are
the corresponding probabilities of measurement outcomes $\{x_{i}\}$.
According to the classical Cramér-Rao inequality, the ultimate precision
of the parameter $\varphi$ is bounded by its classical Fisher information
(CFI) \citep{1925Fisher_information},
\begin{equation}
\delta\varphi\geq\frac{1}{\sqrt{N\mathrm{F}_{\varphi}}},\label{eq:CFI_1}
\end{equation}
where $\delta\varphi$ denotes the stand error of the parameter $\varphi$,
$N$ is the repeated measurement times. The classical Fisher information
is
\begin{equation}
\mathrm{F}_{\varphi}\equiv\sum_{i}p_{i}(\varphi)\left[\frac{\partial\ln p_{i}(\varphi)}{\partial\varphi}\right]^{2}.
\end{equation}
For a continuous variable, the summation should be replaced by an
integral.

When generalized to the quantum estimation theory, the information
of the parameter $\varphi$ is encoded in the density matrix $\rho_{\varphi}$.
The stand error of $\varphi$ is limited by the quantum Fisher information
(QFI),
\begin{equation}
\delta\varphi\geq\frac{1}{\sqrt{N\mathcal{F}_{\varphi}}}.\label{eq:QFI_1}
\end{equation}
The QFI is defined as
\begin{equation}
\mathcal{F}_{\varphi}\equiv\mathrm{Tr}\left(\rho_{\varphi}\hat{L}_{\varphi}^{2}\right),\label{eq:QFI_2}
\end{equation}
where the symmetric logarithmic derivative (SLD) operator $\hat{L}_{\varphi}$
is determined by the derivative of the density matrix $\partial_{\varphi}\rho_{\varphi}\equiv\partial\rho_{\varphi}/\partial\varphi=(\rho_{\varphi}\hat{L}_{\varphi}+\hat{L}_{\varphi}\rho_{\varphi})/2$.
With the spectrum decomposition of the density matrix, i.e., $\rho_{\varphi}=\sum_{i}p_{i}\left|\psi_{i}\right\rangle \left\langle \psi_{i}\right|$,
the QFI becomes
\begin{align}
\mathcal{F}_{\varphi} & =\sum_{i,p_{i}\neq0}\frac{(\partial_{\varphi}p_{i})^{2}}{p_{i}}\nonumber \\
 & +2\sum_{i\neq j,p_{i}+p_{j}\neq0}\frac{(p_{i}-p_{j})^{2}}{p_{i}+p_{j}}\left|\left\langle \psi_{i}|\partial_{\varphi}\psi_{j}\right\rangle \right|^{2}.
\end{align}
 In the above expression, the first term corresponds to the CFI, while
the second term (non-negative) stems from the quantum feature of the
system. Specially, for a two-level quantum system, the QFI is directly
related to its Bloch vector. In the Bloch representation, any two-dimensional
density matrix can be expressed using a real Bloch vector
\begin{equation}
\rho=\frac{1}{2}(\hat{\mathrm{I}}+\boldsymbol{r}\cdot\hat{\boldsymbol{\sigma}}),
\end{equation}
where $\hat{\mathrm{I}}$ stands for the unity matrix, $\hat{\boldsymbol{\sigma}}=(\hat{\sigma}_{x},\hat{\sigma}_{y},\hat{\sigma}_{z})$
characterizes the Pauli matrices, and $\boldsymbol{r}=(r_{x},r_{y},r_{z})^{T}$
denotes the Bloch vector whose modulus satisfies $r\equiv\left|\boldsymbol{r}\right|\leq1$.
The QFI of this density matrix can be cast as \citep{2013zhongwei}

\begin{equation}
\mathcal{F}_{\varphi}=\begin{cases}
\left|\partial_{\varphi}\boldsymbol{r}\right|^{2}+\frac{(\boldsymbol{r}\cdot\partial_{\varphi}\boldsymbol{r})^{2}}{1-r^{2}}, & r<1,\\
\left|\partial_{\varphi}\boldsymbol{r}\right|^{2}, & r=1.
\end{cases}\label{eq:QFI_Bloch}
\end{equation}

\section{Dynamical Modulation \label{sec:DDT-dynamics}}

Here we first unveil the underlying physics of the DM in a quantum
system and then illustrate the relation between the modulation and
the system behavior. We take a two-level system (TLS) that interacts
with a zero-temperature bath as a representative example. A time-varying
control field is introduced to couple with the TLS. The total Hamiltonian
reads ($\hbar=1$)

\begin{align}
\hat{H} & =\frac{\omega_{0}}{2}\hat{\sigma}_{z}+\sum_{k}\omega_{k}\hat{b}_{k}^{\dagger}\hat{b}_{k}+\sum_{k}(g_{k}\hat{\sigma}_{+}\hat{b}_{k}+g_{k}^{*}\hat{b}_{k}^{\dagger}\hat{\sigma}_{-})\nonumber \\
 & +\frac{\xi\nu}{2}\cos(\nu t)\hat{\sigma}_{z}.\label{eq:total_Hamiltonian}
\end{align}
The first two terms characterize the free Hamiltonian of the TLS and
bath where $\hat{\sigma}$ $(\hat{b}_{k})$ stands for the Pauli operator
of the TLS with frequency $\omega_{0}$ (the annihilation operator
of the $k$th bath mode with frequency $\omega_{k}$). The third term
represents the interaction between the TLS and the bath with the coupling
strength $g_{k}$. The last term describes the control field which
modulates the frequency of the TLS periodically, i.e., $\omega_{0}\rightarrow\omega_{0}+\xi\nu\cos(\nu t)$.
Hereinafter we call $\nu$ the modulation frequency and $\xi$ the
modulation amplitude. Clearly, without the control field, the TLS
initially in a superposition of its excited and ground states will
decay to its ground state with vanishing off-diagonal elements in
its density matrix (decoherence).

\begin{figure}[t]
\begin{centering}
\includegraphics[scale=0.35]{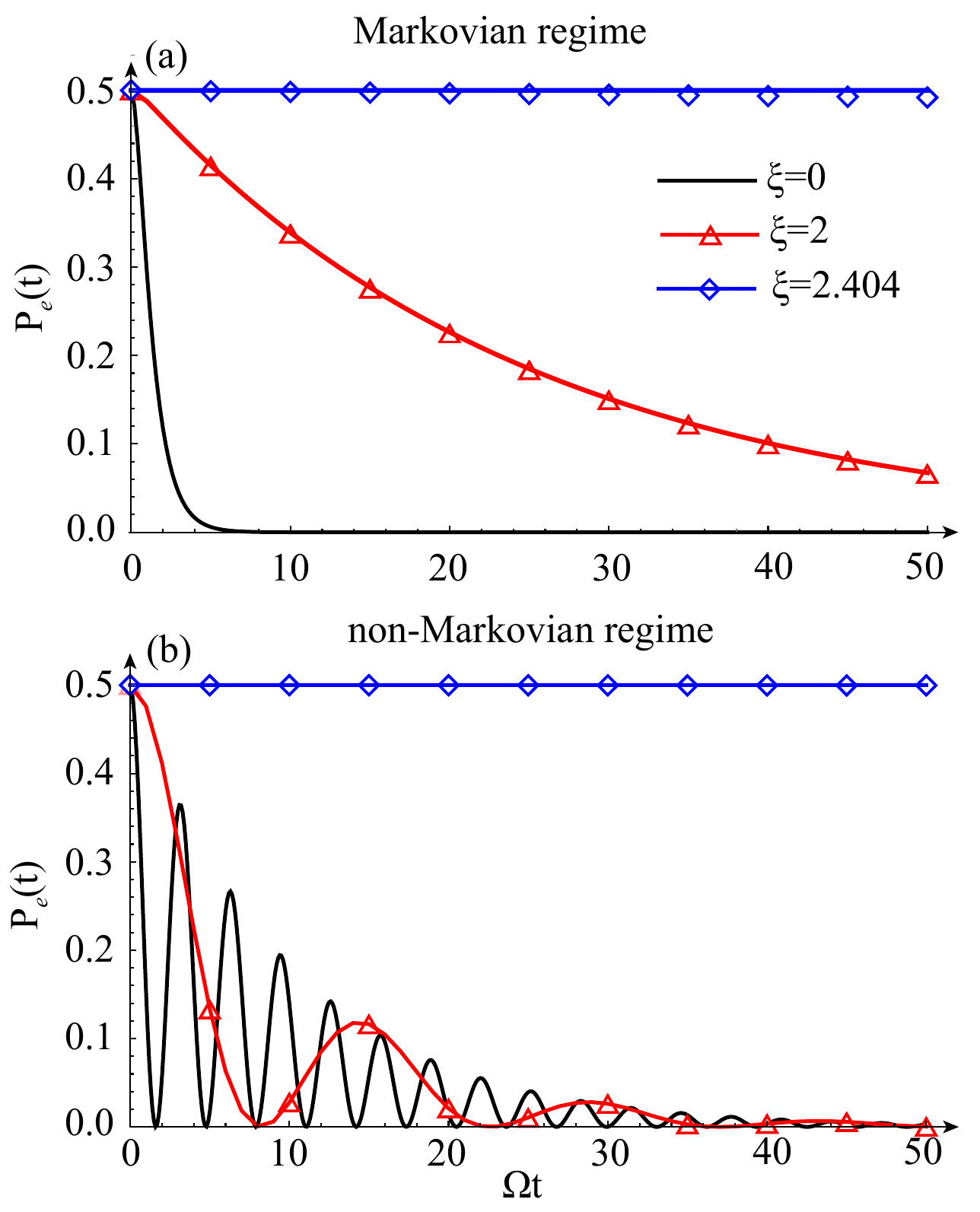}
\par\end{centering}
\caption{\label{fig:probability 1}The excited-state probability $\mathrm{P}_{e}(t)$
of the TLS versus time for different modulation amplitudes $\xi$.
(a) Markovian: $\lambda=5\Omega$. (b) non-Markovian: $\lambda=0.2\Omega$.
Here the initial state is chosen as $\left|\varphi_{s}(0)\right\rangle =(\left|e\right\rangle +\left|g\right\rangle )/\sqrt{2}$.
The black lines show the results without the modulation ($\xi=0$).
The red (blue) lines depict the analytical approximate results (Eq.
(\ref{eq:cet_analy_approx_solution})) for the modulation amplitude
$\xi=2$ $(\xi=2.404)$ while the red triangles (blue diamonds) represent
the corresponding numerical results. Here $\delta_{c}=0$ and $\nu=100\Omega$.}
\end{figure}

\begin{figure}[t]
\begin{centering}
\includegraphics[scale=0.35]{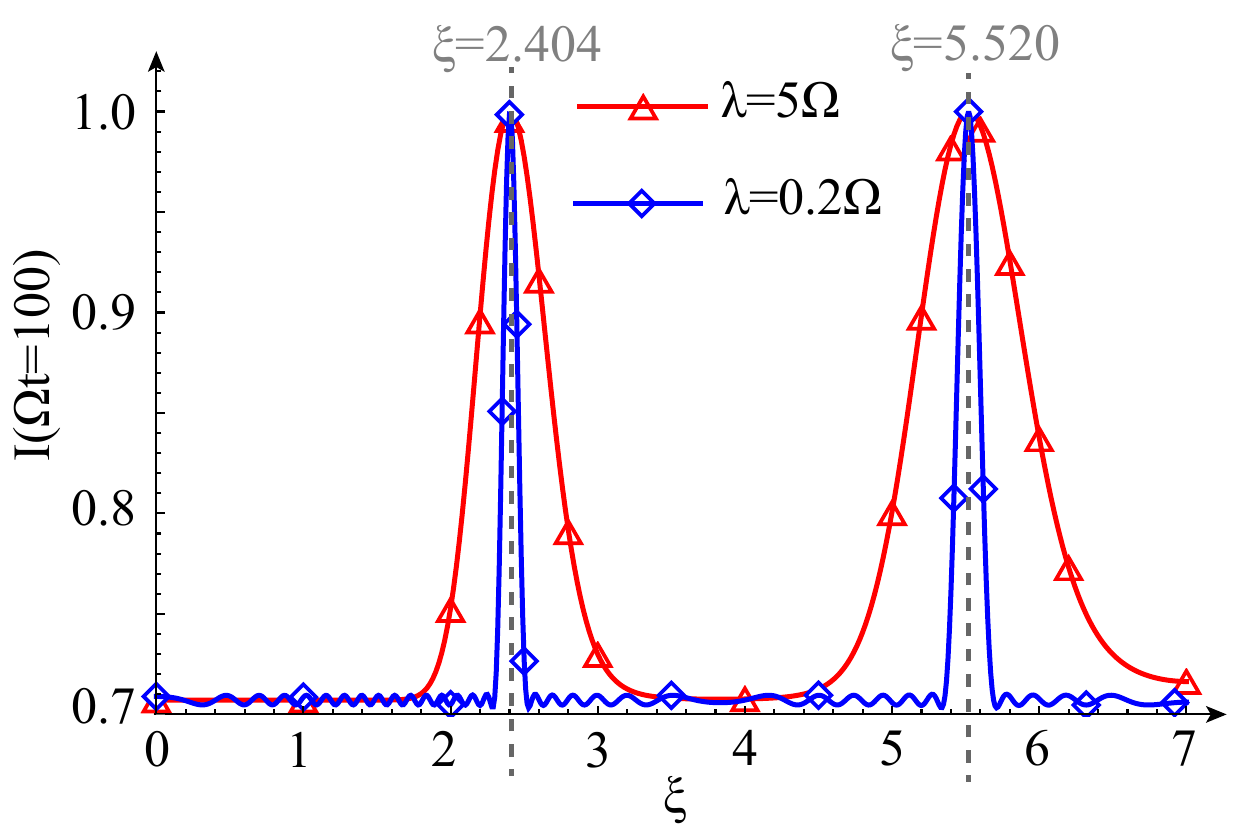}
\par\end{centering}
\caption{\label{fig:fidelity}The fidelity $\mathrm{I}(t)$ of the TLS versus
the modulation amplitude $\xi$. The red (blue) line depicts the analytical
approximate results for $\lambda=5\Omega$ ($\lambda=0.2\Omega$)
with the numerical calculation shown in the red triangles (blue diamonds).
Here $\delta_{c}=0$, $\Omega t=100$, and $\nu=100\Omega$. $\xi=2.404,5.520$
are the zeros of $J_{0}(\xi)$.}
\end{figure}

In the interaction picture with respect to $\hat{H}_{0}=\omega_{0}\hat{\sigma}_{z}/2+\sum_{k}\omega_{k}\hat{b}_{k}^{\dagger}\hat{b}_{k}+\xi\nu\cos(\nu t)\hat{\sigma}_{z}/2$,
the Hamiltonian becomes
\begin{equation}
\hat{H}_{I}=\sum_{k}g_{k}\hat{\sigma}_{+}\hat{b}_{k}e^{i(\omega_{0}-\omega_{k})t+i\xi\sin(\nu t)}+h.c..\label{eq:Interaction Hamil}
\end{equation}
Without loss of generality, we take the initial state as
\begin{equation}
\left|\varphi(0)\right\rangle =\left[c_{g}(0)\left|g\right\rangle +c_{e}(0)\left|e\right\rangle \right]\left|0\right\rangle ,
\end{equation}
where $\left|e\right\rangle (\left|g\right\rangle )$ denotes the
excited (ground) state of the TLS with probability amplitude $c_{e}(0)$
$(c_{g}(0))$. $\left|0\right\rangle \equiv\prod_{k}\left|0_{k}\right\rangle $
describes the vacuum state of the bath. Due to the TLS-bath coupling,
the TLS in the excited state may decay to its ground state by losing
energy to the bath. Thus the state at time $t$ can be written as
\begin{equation}
\left|\varphi(t)\right\rangle =\left[c_{g}(t)\left|g\right\rangle +c_{e}(t)\left|e\right\rangle \right]\left|0\right\rangle +\sum_{k}c_{k}(t)\left|g\right\rangle \left|1_{k}\right\rangle ,\label{eq:phi_t}
\end{equation}
where $\left|1_{k}\right\rangle \equiv\left|1_{k}\right\rangle \otimes\prod_{k^{'}\neq k}\left|0_{k^{'}}\right\rangle $
represents the bath state with only one excitation in the $k$th mode.
As the total excitation is conserved, the amplitude in the zero-excitation
subspace does not vary with time, i.e., $c_{g}(t)=c_{g}(0)$.

The probability amplitude $c_{e}(t)$ satisfies the integro-differential
equation \citep{2001spectral_functiom}
\begin{align}
\dot{c}_{e}(t) & =-\int_{0}^{\infty}d\omega J(\omega)\nonumber \\
 & \times\int_{0}^{t}d\tau c_{e}(\tau)e^{i(\omega_{0}-\omega)(t-\tau)}e^{i\xi\left[\sin(\nu t)-\sin(\nu\tau)\right]},\label{eq:integro-differential_equation}
\end{align}
where we have replaced the summation over the bath modes with an integral
over the frequency in the continuum limit, and $J(\omega)\equiv\sum_{k}\left|g_{k}\right|^{2}\delta(\omega-\omega_{k})$
is the spectral density of the bath. In this work, we consider the
Lorentzian spectral density $J(\omega)=\Omega^{2}\lambda/\left\{ 2\pi\left[(\omega-\omega_{c})^{2}+(\lambda/2)^{2}\right]\right\} $
where $\Omega$ $(\omega_{c})$ represents the coupling strength (central
frequency) of the spectrum. $\lambda$ is the full width at half maximum
of the spectrum whose inverse characterizes the correlation time of
the bath.

The behavior of the system heavily depends on the structure of the
spectral density $J(\omega)$. On the one hand, when the spectral
density is rather ``flat'' in the frequency domain ($\Omega\ll\lambda$,
the weak-coupling regime), the correlation time of the bath is relatively
short. Thus the dynamics of the system at time $t+dt$ is determined
only by that at time $t$, i.e., the memory effect is ignored. In
this case, the dynamics of the system is called ``Markovian''. On
the other hand, when the spectral density is ``structured'' ($\Omega\gg\lambda$,
the strong-coupling regime), the memory effect can not be neglected
anymore. That is, the information of the system at earlier times is
stored in the bath and then flows back to the system \citep{2010QFI_flow_xmlu}.
Such dynamics of the system is called ``non-Markovian''.

With the Jacobi-Anger identity $\exp\left[i\xi\sin(\nu t)\right]=\sum_{n=-\infty}^{\infty}J_{n}(\xi)\exp(in\nu t)$
where $J_{n}(\xi)$ is the $n$th first kind Bessel function, the
solution of the Eq. (\ref{eq:integro-differential_equation}) can
be approximated as (see Appendix \ref{sec:integro-differential equation})

\begin{align}
c_{e}(t) & \simeq c_{e}(0)e^{-(\frac{\lambda}{2}-i\delta_{n_{0}})\frac{t}{2}}\nonumber \\
 & \times\left(\cosh\frac{\varTheta_{n_{0}}t}{2}+\frac{\frac{\lambda}{2}-i\delta_{n_{0}}}{\varTheta_{n_{0}}}\sinh\frac{\varTheta_{n_{0}}t}{2}\right),\label{eq:cet_analy_approx_solution}
\end{align}
where we have defined the notations $\delta_{n}\equiv\delta_{c}+n\nu$,
$\delta_{c}\equiv\omega_{0}-\omega_{c}$, $n_{0}\equiv\left\{ n|\min\left\{ \left|\delta_{n}\right|,n\in Z\right\} \right\} $,
and $\varTheta_{n_{0}}\equiv\sqrt{(\lambda/2-i\delta_{n_{0}})^{2}-4\Omega^{2}J_{n_{0}}^{2}(\xi)}$.
It is worth mentioning that the approximate result Eq. (\ref{eq:cet_analy_approx_solution})
can recover the exact solution of the case without control ($\xi=0$)
\citep{2015phaseestimation} by making the replacement $J_{n_{0}}^{2}(\xi)\rightarrow J_{0}^{2}(0)=1$
and $\delta_{n_{0}}\rightarrow\delta_{0}=\delta_{c}$, which in turn
verifies our derivation here.

Based on the solutions with and without modulation, two features of
the DM emerge: (i) the system-bath coupling strength shrinks from
$\Omega$ to $J_{n_{0}}(\xi)\Omega$ ($\left|J_{n}(\xi)\right|\leq1$
for all $n$); (ii) the frequency of the TLS is shifted from $\omega_{0}$
to $\omega_{0}+n_{0}\nu$. Note that when $\xi$ is set as one of
the zeros of $J_{n_{0}}(\xi)$, the TLS is decoupled from the bath.
In this case, no decoherence occurs, i.e., the TLS is ``frozen''
in its initial state ($c_{e}(t)\simeq c_{e}(0)$).

To show the modulation-induced shrinkage of the coupling strength,
Fig. \ref{fig:probability 1} plots the excited-state probability
$\mathrm{P}_{e}(t)$ of the TLS versus time for different modulation
amplitudes $\xi$. The initial state is chosen as the maximally coherent
state. Apparently, without the control field, the interaction between
the TLS and the bath will cause the decoherence of the TLS (black
lines). When the modulation frequency is larger than the inverse of
the memory time of the environment, the coupling strength is reduced
by the Bessel function (Eq. (\ref{eq:cet_analy_approx_solution})).
As a result, the decoherence of the system is mitigated (red and blue
lines). Notably, when $\xi$ is the zero of the Bessel function ($\xi=2.404$
for our case), the decoherence of the system is completely suppressed
by dynamically decoupled from the noise. We also illustrate the fidelity
$\mathrm{I}(t)\equiv\mathrm{Tr}\sqrt{\rho_{s}^{1/2}(0)\rho_{s}(t)\rho_{s}^{1/2}(0)}$
of the system in Fig. \ref{fig:fidelity} which depicts the deviation
of the state from its initial state. In the steady state, the fidelity
$\mathrm{I}(\Omega t=100)$ remains almost unity at the zeros of the
Bessel function $J_{0}(\xi)$ ($n_{0}=0$ in our case, see the vertical
gray dashed lines) and decreases to $1/\sqrt{2}$ rapidly for other
cases. Our analysis is valid for both the Markovian and non-Markovian
baths, demonstrated by the numerical calculations (the triangles and
diamonds).

\begin{figure}[t]
\begin{centering}
\includegraphics[scale=0.35]{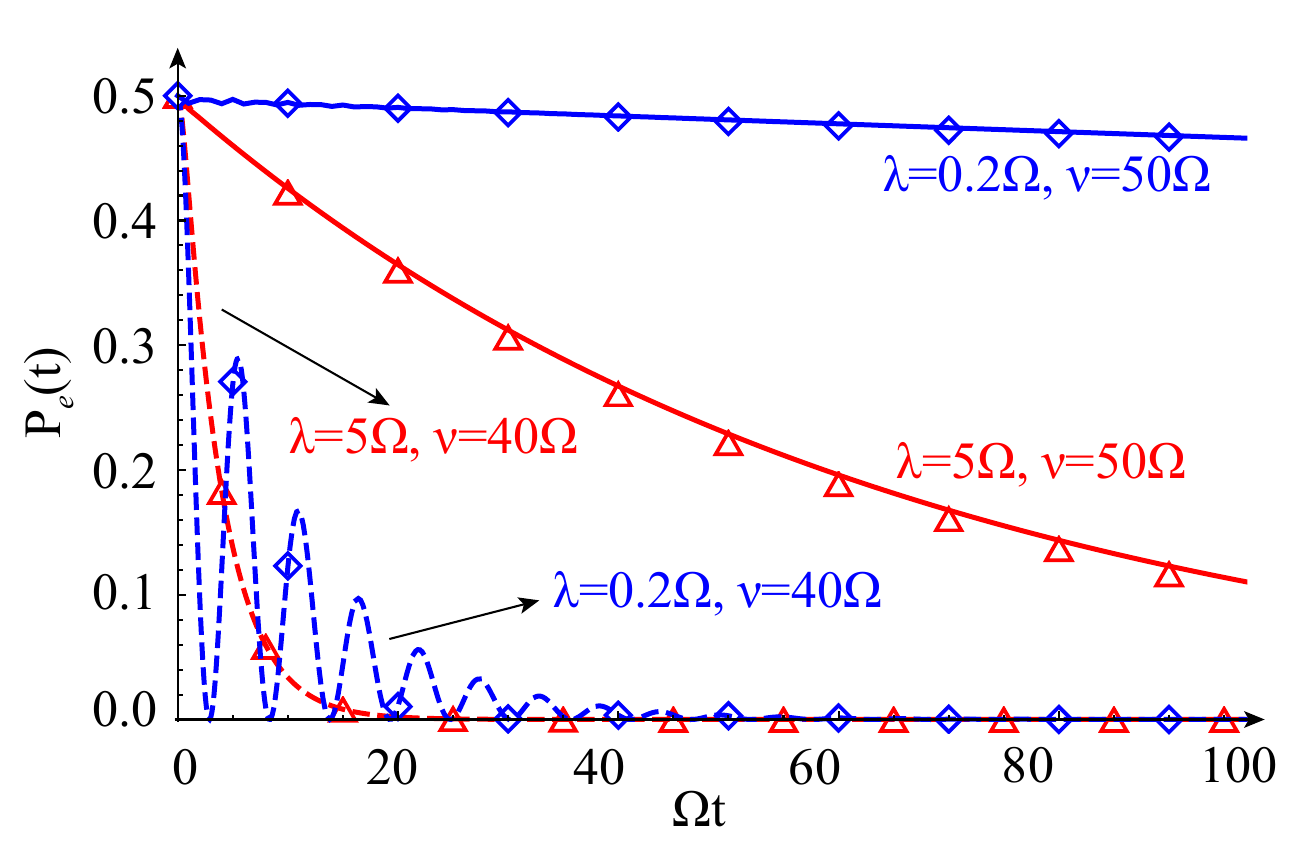}
\par\end{centering}
\caption{\label{fig:probability 2}The excited-state probability $\mathrm{P}_{e}(t)$
of the TLS versus time for different modulation frequencies $\nu$.
The red and blue results stand for $\mathrm{P}_{e}(t)$ in the Markovian
($\lambda=5\Omega$) and non-Markovian ($\lambda=0.2\Omega$) regimes.
The solid (dashed) lines plot the analytical approximate results for
the modulation frequency $\nu=50\Omega$ ($\nu=40\Omega$) with the
numerical results shown in the triangles and diamonds. Here $\delta_{c}=40\Omega$
and $\xi=2$.}
\end{figure}

Figure \ref{fig:probability 2} displays another feature of the modulation,
i.e., the shift of the system's effective frequency. It is remarkable
that besides the coupling-strength reduction, the frequency shift
can also modulate the dynamics of the system especially when the amplitude
$\xi$ is not exactly the zero of the Bessel function. In other words,
the shift of the effective frequency will change the response of the
TLS to the bath, for example, the stationary distribution and the
decay rate. In Fig. \ref{fig:probability 2}, we choose $\delta_{c}\equiv\omega_{0}-\omega_{c}=40\Omega$
and alter the modulation frequency from $\nu=40\Omega$ to $\nu=50\Omega$
(the effective frequency from $\omega_{c}$ to $\omega_{c}-10\Omega$).
As is shown, the behavior of the probability differs from each other
due to this frequency shift, which coincides with our approximate
analysis in Eq. (\ref{eq:cet_analy_approx_solution}). Since we consider
the bath at zero temperature, the atom decays to its ground state
in the long-time limit ($\mathrm{P}_{e}(t)\sim0$). For a thermal
bath (non-zero temperature), the frequency shift will lead to a weighted
superposition of the excited-state probability (see the discussion
in Sec. \ref{subsec:Quantum-thermometry}).

\section{\label{sec:enhanced-parameter-estimatio}DM-enhanced parameter-estimation
precision}

In this section, we concentrate on the application of the DM in the
quantum parameter estimation. With several illustrative examples,
we reveal that the precision promotion in two estimation types stems
from the two features of the DM respectively. When the unknown parameter
is independent of the bath (the Ramsey spectroscopy as an example),
the DM enhances the precision by decoherence suppression. When the
parameter arises from the bath (the quantum thermometer as an example),
the DM improves the estimation performance through the effective frequency
selecting. To be more specific, since the dynamics of the system varies
with its frequency and is sensitive to the bath, a careful selection
of the system's effective frequency will boost the information flow
from the bath to the system (see the following for details).

\subsection{Type I: Ramsey Spectroscopy\label{subsec:Ramsey_spectroscopy}}

In a Ramsey spectroscopy, the frequency $\omega_{0}$ between two
atomic levels is the parameter of interest. The corresponding parametrization
process is achieved by three steps. At first, the atom in its ground
state evolves to the maximally coherent state $(\left|g\right\rangle -i\left|e\right\rangle )/\sqrt{2}$
by a $\pi/2$ pulse. Then, the atom experiences a free evolution with
time $T_{f}$, encoding the frequency information into the relative
phase of the atom. Finally, after a second $\pi/2$ pulse, the frequency
information is stored in the excited- and ground-state probabilities
of the atom. In an ideal case (without noise), the projection measurement
$\left\{ \left|g\right\rangle \left\langle g\right|,\left|e\right\rangle \left\langle e\right|\right\} $
saturates the quantum Cramér-Rao bound with the QFI $\mathcal{F}_{\omega_{0}}(T_{f})=T_{f}^{2}$
\citep{1997atomic_spectroscopy}.

In contrast, for a noisy Ramsey spectroscopy, the system-environment
interaction induces the decoherence of the system and thus degrades
the information in the relative phase. To prevent such information
leakage, we apply the DM to manipulate the system's dynamics. Since
the free-evolution time $T_{f}$ is much larger than the pulse duration
time, here we ignore the system decoherence in the pulse duration
and investigate the performance of the DM in the free-evolution process.

With the time-varying field, the total Hamiltonian is the same as
that in Eq. (\ref{eq:total_Hamiltonian}). After the free evolution,
the state of the whole system can be expressed as (in the Schrödinger
picture)
\begin{align}
 & \left|\varphi(T_{f})\right\rangle \nonumber \\
 & =e^{\frac{i}{2}\Xi(\omega_{0},T_{f})}\left[c_{g}(T_{f})\left|g\right\rangle +e^{-i\Xi(\omega_{0},T_{f})}c_{e}(T_{f})\left|e\right\rangle \right]\left|0\right\rangle \nonumber \\
 & +\sum_{k}e^{-i\omega_{k}T_{f}}c_{k}(T_{f})\left|g\right\rangle \left|1_{k}\right\rangle ,\label{eq:state_after_free_evolution}
\end{align}
where we have defined the phase $\Xi(\omega_{0},T_{f})\equiv\omega_{0}T_{f}+\xi\sin(\nu T_{f})$.
The initial conditions are $c_{g}(0)=1/\sqrt{2}$, $c_{e}(0)=-i/\sqrt{2}$,
and $c_{k}(0)=0$. The dynamical equation of $c_{e}(T_{f})$ is shown
in Eq. (\ref{eq:integro-differential_equation}) with its analytical
approximate solution given in Eq. (\ref{eq:cet_analy_approx_solution}).

According to the quantum Cramér-Rao inequality, the estimation precision
of the parameter $\omega_{0}$ is ultimately limited by its QFI ($\left|R(T_{f})\right|\neq0,1$)

\begin{align}
\mathcal{F}_{\omega_{0}}(T_{f}) & =T_{f}^{2}\left|R(T_{f})\right|^{2}+\frac{\left[\frac{\partial\left|R(T_{f})\right|}{\partial\omega_{0}}\right]^{2}}{1-\left|R(T_{f})\right|^{2}}\nonumber \\
 & +\left|\frac{\partial R(T_{f})}{\partial\omega_{0}}\right|^{2}+2T_{f}\mathrm{Im}\left[R(T_{f})\frac{\partial R^{*}(T_{f})}{\partial\omega_{0}}\right],\label{eq:QFI_Ramsey}
\end{align}
where we have denoted the ratio by $R(T_{f})\equiv c_{e}(T_{f})/c_{e}(0)$.
$\left|R(T_{f})\right|=1~(0)$ corresponds to the pure-state case
where the parameter information is completely retained in the system
$\mathcal{F}_{\omega_{0}}(T_{f})=T_{f}^{2}$ (lost to the bath $\mathcal{F}_{\omega_{0}}(T_{f})=0$).
Clearly, apart from the excited-state probability of the atom, the
QFI also depends on the derivative of excited-state probability amplitude
with respect to $\omega_{0}$, i.e., $\partial R(T_{f})/\partial\omega_{0}$,
which renders Eq. (\ref{eq:QFI_Ramsey}) rather cumbersome to exhibit
the impact of the DM visually. Further investigation implies that
for a high-frequency modulation field, the derivative $\partial R(T_{f})/\partial\omega_{0}$
vanishes (see Appendix \ref{sec:Ramsey spectroscopy}). Thus, the
analytical approximation of $\mathcal{F}_{\omega_{0}}(T_{f})$ can
be cast as

\begin{align}
\mathcal{F}_{\omega_{0}}(T_{f}) & \simeq T_{f}^{2}\left|R(T_{f})\right|^{2}.\label{eq:QFI_Ramsey-appro}
\end{align}
Particularly, one can retrieve the pure-state QFI as $\mathcal{F}_{\omega_{0}}(T_{f})=T_{f}^{2}~(0)$
by making the replacement $\left|R(T_{f})\right|\rightarrow1~(0)$
in Eq. (\ref{eq:QFI_Ramsey-appro}).

\begin{figure}[t]
\begin{centering}
\includegraphics[scale=0.35]{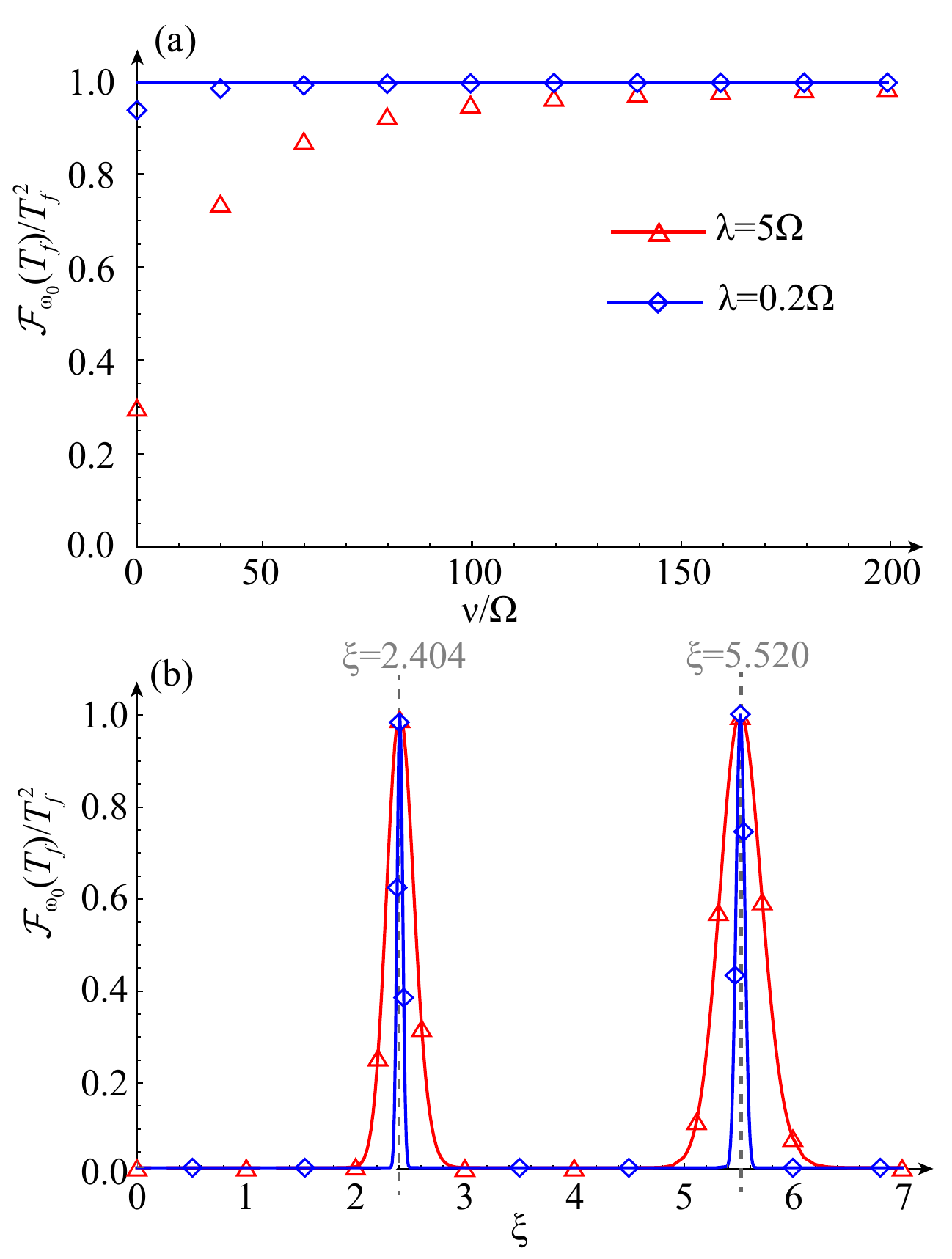}
\par\end{centering}
\caption{\label{fig:QFI_Ramsey}The normalized QFI $\mathcal{F}_{\omega_{0}}(T_{f})/T_{f}^{2}$
versus the modulation frequency $\nu$ (a) and amplitude $\xi$ (b).
The red (blue) results stand for the QFI in the Markovian (non-Markovian)
regime. The solid lines show the analytical approximate results (Eq.
(\ref{eq:QFI_Ramsey-appro})) while the triangles and diamonds give
the numerical calculations (Eq. (\ref{eq:QFI_Ramsey})). Here $\xi=2.404$
in (a), $\nu=200\Omega$ in (b), $\delta_{c}=0$, and $\Omega T_{f}=150$.}
\end{figure}

Equation (\ref{eq:QFI_Ramsey-appro}) verifies the decoherence-suppression
promotion of the QFI in the Ramsey spectroscopy. As shown in Eqs.
(\ref{eq:cet_analy_approx_solution}) and (\ref{eq:QFI_Ramsey-appro}),
the shrunk coupling strength $\Omega J_{n_{0}}(\xi)$ mitigates the
decay of the ratio $\left|R(T_{f})\right|$ and protects the information
from noise, resulting in a large QFI. As an extreme case, when $\xi$
is chosen as one of the zeros of $J_{n_{0}}(\xi)$, the decoherence
of the system is suppressed entirely ($R(T_{f})\simeq1$), and the
QFI recovers that of the ideal case, i.e., $\mathcal{F}_{\omega_{0}}(T_{f})\simeq T_{f}^{2}$.
Fig. \ref{fig:QFI_Ramsey} plots the variation of the normalized QFI
$\mathcal{F}_{\omega_{0}}(T_{f})/T_{f}^{2}$ versus the modulation
frequency $\nu$ and amplitude $\xi$, which demonstrates the coincidence
between our approximate analysis (solid lines, Eq. (\ref{eq:QFI_Ramsey-appro}))
and numerical calculation (triangles and diamonds, Eq. (\ref{eq:QFI_Ramsey})).

\begin{figure}[t]
\begin{centering}
\includegraphics[scale=0.17]{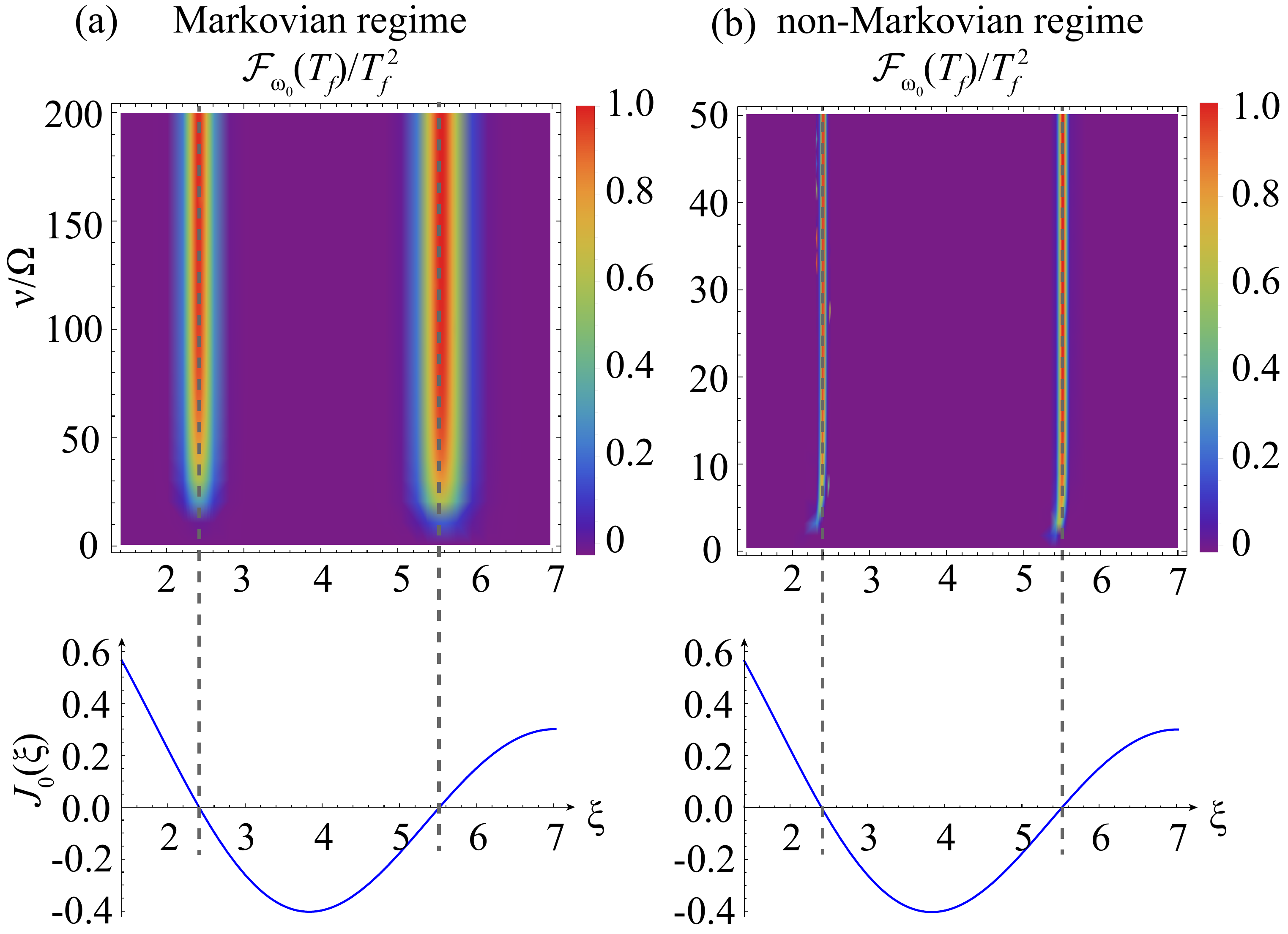}
\par\end{centering}
\caption{\label{fig:2D_QFI_Ramsey}The contours of the normalized QFI $\mathcal{F}_{\omega_{0}}(T_{f})/T_{f}^{2}$
under the Markovian ($\lambda=5\Omega$) (a) and the non-Markovian
($\lambda=0.2\Omega$) dynamics (b) versus the modulation amplitude
$\xi$ and frequency $\nu$. The vertical gray dashed lines $\xi=2.404,5.520$
represent the zeros of $J_{0}(\xi)$. Here $\delta_{c}=0$ and $\Omega T_{f}=150$.}
\end{figure}

To show the influence of the DM intuitively, we present the contour
plots of the normalized QFI $\mathcal{F}_{\omega_{0}}(T_{f})/T_{f}^{2}$
versus the modulation frequency and amplitude in Fig. \ref{fig:2D_QFI_Ramsey}.
When the modulation frequency is larger than the inverse of the reservoir's
memory time, the QFI regains that of the ideal case at exactly the
zeros of the Bessel function and vanishes rapidly in other cases.
Our DM-enhanced parameter estimation proposal exhibits a good performance
in both the Markovian and non-Markovian regimes.

\subsection{\label{subsec:Quantum-thermometry}Type II: Quantum Thermometer}

In the above subsection, we have illustrated that the DM can retain
the information in the system by decoherence suppression in the estimation
type I. In type II, since the system-bath interaction is a prerequisite
for the parametrization process, the validity of the precision enhancement
using the DM remains elusive. In the following, we take the quantum
thermometer as a representative example and elucidate the precision
improvement arising from the effective frequency selection.

\subsubsection{The conventional quantum thermometer ($\xi=0$)}

Before investigating the DM-assisted quantum thermometer, we first
revisit the working mechanism of the conventional quantum thermometer
(without the modulation) and present the lower bound of its temperature
uncertainty. Here we consider the smallest possible thermometer, i.e.,
a TLS, which minimizes the perturbation on the bath (sample). In the
long-time limit, the TLS will be thermalized to the bath temperature
$T$ with the quantum Fisher information as ($k_{b}=1$) \citep{2015thermometry,2018Campbell_thermometer,2018thermometry}
\begin{equation}
\mathcal{F}_{T}^{0}(\omega_{0})=\frac{\omega_{0}^{2}}{4T^{4}\cosh^{2}(\omega_{0}/2T)},\label{eq:FI_thermo no modulation}
\end{equation}
where the superscript ``0'' denotes the case without the DM. Further
analysis indicates that for a given TLS (fixed $\omega_{0}$), the
QFI first increases and then decreases as the temperature grows (single
peak) with the optimal temperature $T_{\mathrm{max}}$ satisfying
\begin{equation}
\coth\left(\frac{\omega_{0}}{2T_{\mathrm{max}}}\right)=\frac{\omega_{0}}{4T_{\mathrm{max}}}.\label{eq:optimal_temperature_equation}
\end{equation}
Any deviation from this optimal temperature ($T_{\mathrm{max}}\simeq0.242\omega_{0}$)
will deteriorate the estimation precision. Particularly, the QFI in
Eq. (\ref{eq:FI_thermo no modulation}) diverges in the zero-temperature
limit, which limits the application of the quantum thermometer in
domains such as ultra-cold gases \citep{2020Bouton_ultracoldgas}
and superconductivity \citep{2010Rosenstein_superconductor}.

\subsubsection{DM-enhanced quantum thermometer ($\xi\protect\neq0$)}

In the quantum thermometer with the DM, due to the DM-induced frequency
shift illustrated in Sec. \ref{sec:DDT-dynamics}, more than one peaks
will emerge in the QFI, broadening the efficient operation range of
the thermometer.

\begin{figure}[t]
\begin{centering}
\includegraphics[scale=0.35]{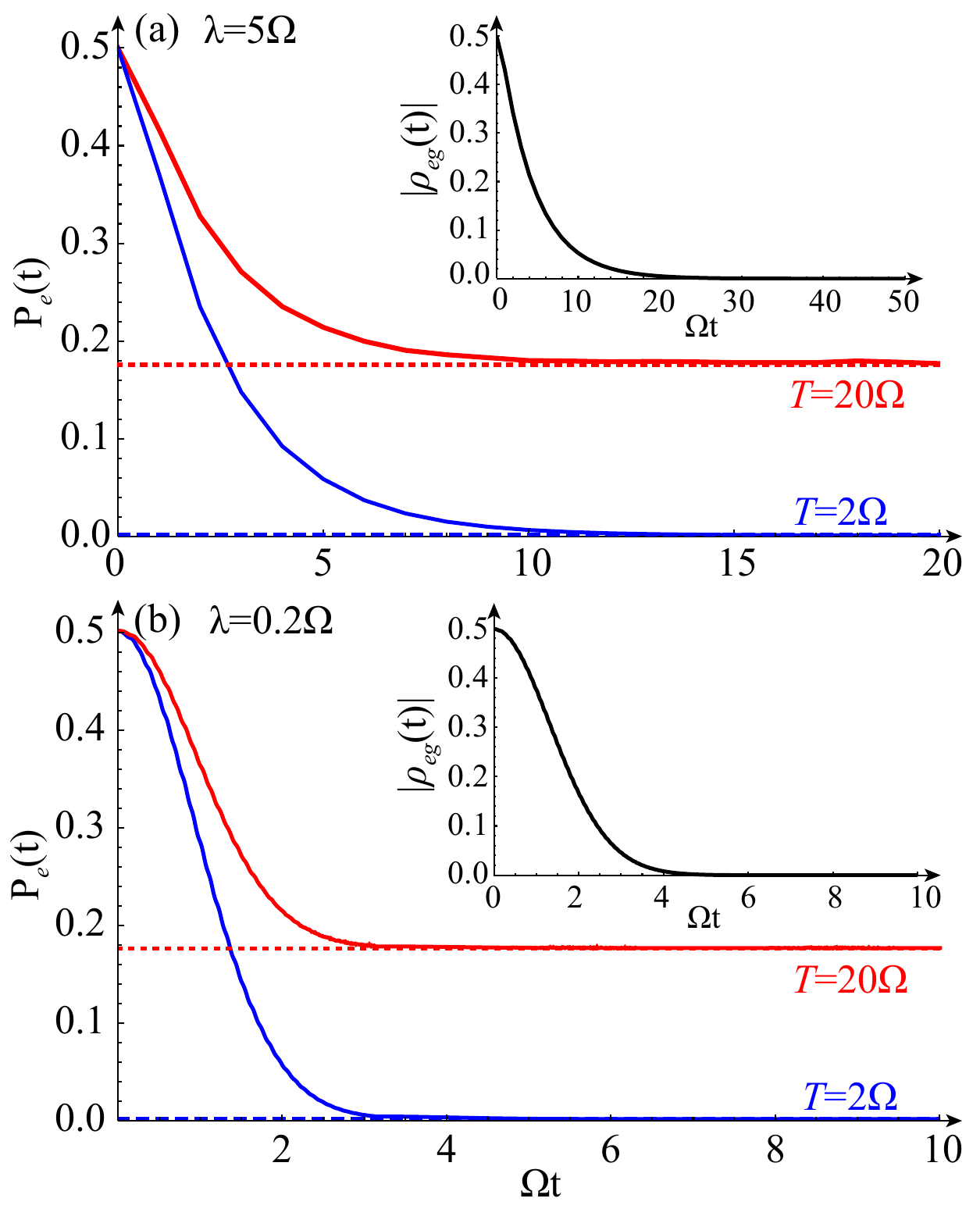}
\par\end{centering}
\caption{\label{fig:Pe eg t}The dynamics of the excited-state probability
$\mathrm{P}_{e}(t)$ under the Markovian ($\lambda=5\Omega$) dynamics
(a) and the non-Markovian ($\lambda=0.2\Omega$) dynamics (b). The
initial state of the TLS is $\left|\varphi_{s}(0)\right\rangle =(\left|e\right\rangle +\left|g\right\rangle )/\sqrt{2}$.
The blue solid (dashed) lines show the numerical (analytical approximate)
results for the case of $T=2\Omega$ while red solid (dotted) lines
plot that for the case of $T=20\Omega$. The insets describe the dynamics
of the off-diagonal term $\left|\rho_{eg}(t)\right|$. Here $\omega_{0}=31\Omega$,
$\delta_{c}=0$, $\nu=30\Omega$, and $\xi=1.0$.}
\end{figure}

The Hamiltonian of the total system in the interaction picture has
been displayed in Eq. (\ref{eq:Interaction Hamil}). Here the second-order
time-convolutionless (TCL) master equation is utilized to analyse
the estimation processes in both the Markovian and non-Markovian regimes
\citep{2002open_system,2010_TCL_NZ,2014exact_TCL_NZ,2019weimin_Zhang},
\begin{equation}
\dot{\rho}_{s}(t)\simeq-\int_{0}^{t}\mathrm{Tr}_{b}\left([\hat{H}_{I}(t),[\hat{H}_{I}(\tau),\rho_{s}(t)\otimes\rho_{b}]]\right)d\tau,
\end{equation}
where the bath is in its thermal state $\rho_{b}\equiv e^{-\hat{H}_{b}/T}/\mathrm{Tr}\left(e^{-\hat{H}_{b}/T}\right)$
with $\hat{H}_{b}\equiv\sum_{k}\omega_{k}\hat{b}_{k}^{\dagger}\hat{b}_{k}$.
In the Markovian regime, this TCL master equation will reduce to the
Markovian master equation \citep{2002open_system}.

Inserting the Hamiltonian into the master equation and taking the
partial trace over the bath, we obtain the dynamical equations of
the diagonal and off-diagonal elements of the density matrix (see
Appendix \ref{sec:appen_TCL_equation} for details)

\begin{align}
\dot{\mathrm{P}}_{e}(t) & \simeq\int_{0}^{t}d\tau\int_{0}^{\infty}d\omega J(\omega)2\cos\left[\phi(t)-\phi(\tau)\right]\nonumber \\
 & \times\left[N_{+}(\omega)-\mathrm{P}_{e}(t)\right],\nonumber \\
\dot{\rho}_{eg}(t) & \simeq-\int_{0}^{t}d\tau\int_{0}^{\infty}d\omega J(\omega)e^{i\left[\phi(t)-\phi(\tau)\right]}\rho_{eg}(t).\label{eq:TCl_equa}
\end{align}
We have defined the phase $\phi(t)\equiv(\omega_{0}-\omega)t+\xi\sin(\nu t)$
and dropped the subscript ``$s$'' for simplicity. Without loss
of generality, we focus on the fermionic-bath case where the mean
particle number is $N_{+}(\omega)\equiv(e^{\omega/T}+1)^{-1}$ (see
Appendix \ref{sec:appen-bosonic-bath} for the bosonic-bath case).

As shown in Sec. \ref{sec:DDT-dynamics}, the frequency of the system
is shifted as a series of discrete frequencies separated by the modulation
frequency. In the long-time limit, the behavior of the TLS can be
regarded as a superposition of systems with equally separated frequencies.
In other words, the excited-state probability in the steady state
is a weighted superposition of a series of equilibrium distributions

\begin{align}
\mathrm{P}_{e} & \equiv\lim_{t\rightarrow\infty}\mathrm{P}_{e}(t)=\sum_{n}P_{n}N_{+}(\omega_{n}).\label{eq:steady state Pe}
\end{align}
The weight of the $n$th component is $P_{n}=J_{n}^{2}(\xi)J(\omega_{n})/\left[\sum_{m}J_{m}^{2}(\xi)J(\omega_{m})\right]$
which stems from the spectral density and the Bessel function. The
coherence of the TLS vanishes in the steady state, i.e., $\lim_{t\rightarrow\infty}\rho_{eg}(t)=0$.

\begin{figure}[t]
\begin{centering}
\includegraphics[scale=0.36]{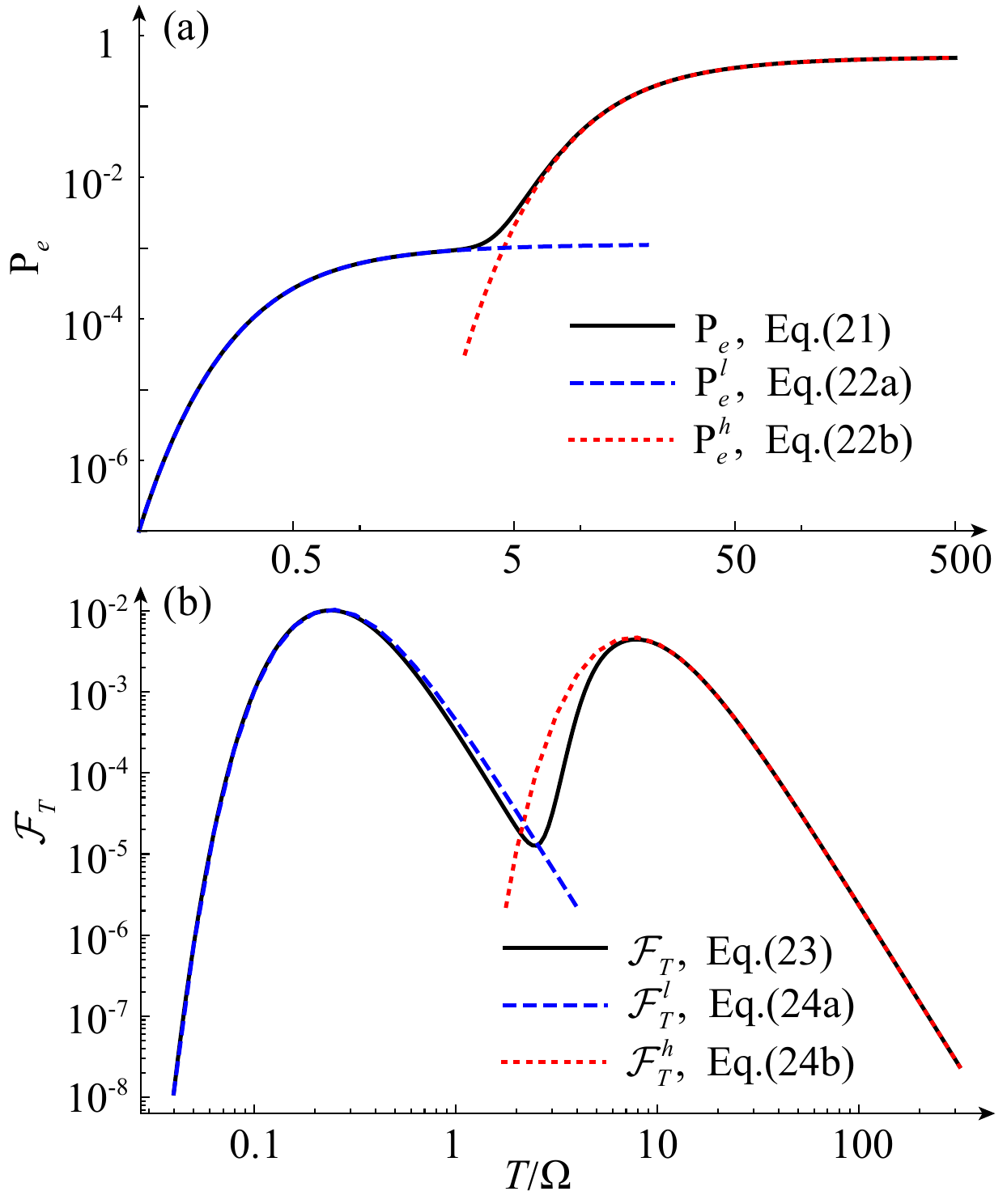}
\par\end{centering}
\caption{\label{fig:Pe=000026FI_tem} The excited-state probability $\mathrm{P}_{e}$
(a) and the QFI $\mathcal{F}_{T}$ (b) versus the temperature. The
black lines plot the analytical results while the blue dashed (red
dotted) lines show the approximation in the low-temperature (high-temperature)
region. The parameters are the same as in Fig. \ref{fig:Pe eg t}(a).}
\end{figure}

Figure \ref{fig:Pe eg t} demonstrates our analysis on the dynamics
of the thermometer. The numerical calculation of the master equation
(solid lines, Eq. (\ref{eq:TCl_equa})) shows that the excited-state
probability $\mathrm{P}_{e}(t)$ first varies with time and then reaches
its steady-state value which is perfectly predicted by our approximate
analysis (dashed lines, Eq. (\ref{eq:steady state Pe})). In the inset,
the vanishing of the modulus of the off-diagonal term $\left|\rho_{eg}(t)\right|$
elucidates the complete decoherence of the TLS. Comparing Figs. \ref{fig:Pe eg t}(a)
with (b), we also validate the usefulness of our proposal in both
the Markovian and non-Markovian regimes.

\begin{figure}[t]
\begin{centering}
\includegraphics[scale=0.35]{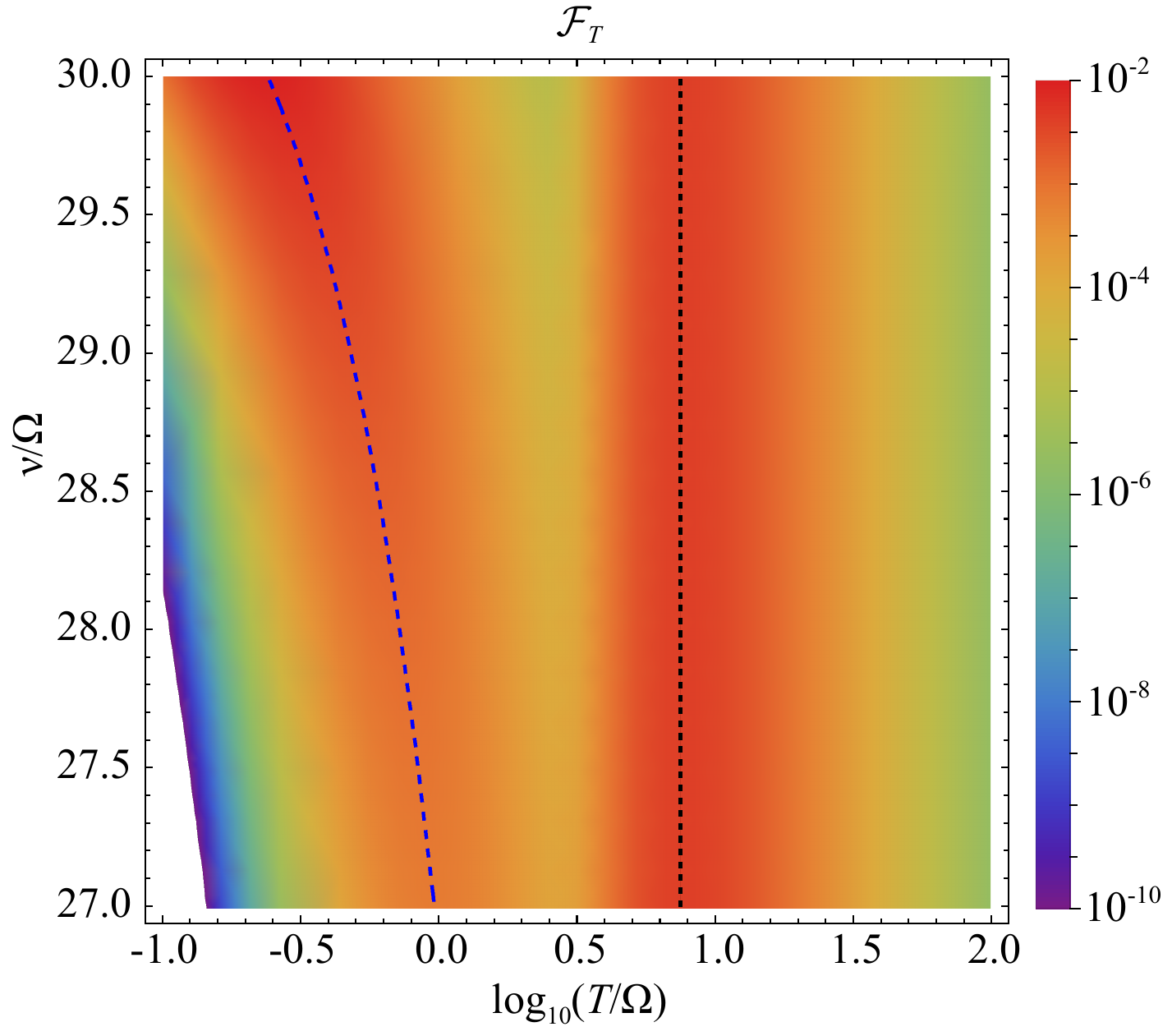}
\par\end{centering}
\caption{\label{fig:2D_FI_thermo}The contour of the QFI $\mathcal{F}_{T}$
as a function of the temperature $T$ and the modulation frequency
$\nu$. The blue dashed and black dotted lines represent the two optimal
temperatures ($0.242\omega_{n_{1}}$ and $0.242\omega_{n_{0}}$).
The parameters are the same as in Fig. \ref{fig:Pe=000026FI_tem}.}
\end{figure}

The superposed probability $\mathrm{P}_{e}$ implies the possibility
of a thermometer with multiple effective frequencies. By defining
the effective frequency as $\omega_{\mathrm{eff}}\equiv T\ln\left(\mathrm{P}_{e}^{-1}-1\right)$
with Eq. (\ref{eq:steady state Pe}), we can analyze the manipulation
of the DM on the effective frequency. For a large modulation frequency
$\nu$, the series of distributions $N_{+}(\omega_{n})$ approximately
form a geometric sequence with the common ratio $e^{-\nu/T}$. In
the low-temperature region, this common ratio is much smaller than
unity. Hence the dominant term in Eq. (\ref{eq:steady state Pe})
is that with the minimum frequency $n_{1}\equiv-\left[\frac{\omega_{0}}{\nu}\right]$
($\left[\circ\right]$ represents the floor function), i.e., $\omega_{\mathrm{eff}}\simeq\omega_{n_{1}}$.
In the high-temperature region, $N_{+}(\omega_{n})$ is rather smooth
with respect to $P_{n}$. $\mathrm{P}_{e}$ is dominated by the term
with the maximum spectrum density, i.e., $\omega_{\mathrm{eff}}\simeq\omega_{n_{0}}$,
$n_{0}\equiv\left\{ n|\mathrm{min}\left\{ \left|\delta_{n}\right|,n\in Z\right\} \right\} $.
In consequence, two effective frequencies of the TLS emerge with the
corresponding steady-state probability as

\begin{subequations}
\begin{align}
\mathrm{P}_{e}^{l} & \simeq P_{n_{1}}N_{+}(\omega_{n_{1}}),\\
\mathrm{P}_{e}^{h} & \simeq N_{+}(\omega_{n_{0}}).\label{eq:Pe h=000026l}
\end{align}
\end{subequations}The superscripts ``$l$'' and ``$h$'' denote
the low- and high-temperature regions. Note also that $n_{1}\leq n_{0}$
and thus $\omega_{n_{1}}\leq\omega_{n_{0}}$.

In the steady state, the QFI of the thermometer with respect to the
temperature reads

\begin{equation}
\mathcal{F}_{T}=\frac{1}{T^{4}}\frac{\left[\frac{1}{4}\sum_{n}P_{n}\omega_{n}\frac{1}{\cosh^{2}\left(\omega_{n}/2T\right)}\right]^{2}}{\mathrm{P}_{e}(1-\mathrm{P}_{e})}.\label{eq:FI_thermo}
\end{equation}
Similarly, the Fisher information in the low- and high-temperature
region can be approximated as \begin{subequations}
\begin{align}
\mathcal{F}_{T}^{l} & \simeq P_{n_{1}}\frac{\omega_{n_{1}}^{2}}{4T^{4}\cosh^{2}(n_{1}/2T)}=P_{n_{1}}\mathcal{F}_{T}^{0}(\omega_{n_{1}}),\label{eq:QFI low}\\
\mathcal{F}_{T}^{h} & \simeq\frac{\omega_{n_{0}}^{2}}{4T^{4}\cosh^{2}(n_{0}/2T)}=\mathcal{F}_{T}^{0}(\omega_{n_{0}}).\label{eq:FI_thermo l=000026h}
\end{align}
\end{subequations}

Equations (\ref{eq:QFI low}) and (\ref{eq:FI_thermo l=000026h})
manifest a double-peak characteristic of the QFI inherited from the
double-frequency feature of $\mathrm{P}_{e}$. As indicated in Eqs.
(\ref{eq:FI_thermo no modulation}) and (\ref{eq:optimal_temperature_equation}),
a precise temperature estimation demands an increase of the system's
frequency with the growth of the temperature, which is not feasible
in the conventional quantum thermometer. With the assistance of modulation,
the QFI in the low- and high-temperature regions behaves distinctly
different. Moreover, since the effective frequency in the low-temperature
region can be lower than that in the high-temperature region, two
peaks appear in the QFI with respect to $\omega_{n_{1}}$ and $\omega_{n_{0}}$,
which significantly broadens the efficient operation range of the
thermometer.

Figures \ref{fig:Pe=000026FI_tem}(a) and \ref{fig:Pe=000026FI_tem}(b)
present the double-frequency and double-peak phenomena of the probability
and QFI, respectively. For simplicity, we only display the results
in the Markovian regime. With the parameters $\nu=30\Omega$, $\omega_{0}=31\Omega$,
and $\delta_{c}=0$, the two effective frequencies are $\omega_{n_{1}}=\omega_{0}-\nu$
and $\omega_{n_{0}}=\omega_{0}$. In Fig. \ref{fig:Pe=000026FI_tem}(a),
$\mathrm{P}_{e}$ coincides with the system with a smaller frequency
in the low-temperature region (the blue dashed line) while remaining
unchanged in the high-temperature region (the red dotted line). Consequently,
a new peak of the Fisher information emerges to the left of the original
one (Fig. \ref{fig:Pe=000026FI_tem}(b)), resulting in the double-peak
characteristics of the QFI.

In Fig. \ref{fig:2D_FI_thermo}, the contour plot of the QFI exhibits
the double-peak phenomenon more explicitly. As predicted by Eqs. (\ref{eq:QFI low})
and (\ref{eq:FI_thermo l=000026h}), the left peak (the emerging one)
increases as the rise of the modulation frequency with its peak position
tending to a lower temperature (the blue dashed line, $T_{\mathrm{max}}\simeq0.242(\omega_{0}-\nu)$).
Meanwhile, the right peak (the original one) maintains its position
and value (the black dotted line, $T_{\mathrm{max}}\simeq0.242\omega_{0}$)
with the variation of the modulation frequency. In this sense, the
DM extends the efficient operation range of the thermometer to the
low-temperature region by promoting its corresponding QFI without
damaging the precision in other regions.

\section{Discussions and conclusions\label{sec:Discussions-and-conclusions}}

In conclusion, we have proposed a general dynamical-modulation-based
quantum parameter estimation scheme that enables the precision enhancement
regardless of the origin of the unknown parameter. Further research
reveals that such improvement originates from the two features of
the DM, i.e., the coupling-strength shrinkage and the effective-frequency
selection. Namely, beyond the capability of decoherence suppression
in the estimation type I, the DM manipulates the dynamics of the system
by frequency selecting, which is sensitive to the bath parameter,
leading to a boost of the information flow to the system in type II.

To be more specific, by taking the Ramsey spectroscopy and quantum
thermometer as typical examples, we investigate the estimation processes
with modulation and uncover the underlying mechanism of the precision
enhancement. In the Ramsey spectroscopy (type I), the leakage of the
phase information to the bath is mitigated by the DM-induced shrinkage
of the system-bath coupling. Specially, when the modulation amplitude
is chosen as the zero of the Bessel function, the system is dynamically
decoupled from the environment, manifesting the fully retention of
the information in the system. As a contrast, in the quantum thermometer
(type II) the temperature information flows from the bath to the system
through the thermalization process. With the DM, the system behaves
as a weighted superposition of a series of systems with frequencies
separated by the modulation frequency. More importantly, for a meticulous
choice of the modulation parameter, the low-temperature effective
frequency is lower than the high-temperature one. Hence, unlike the
single-peak phenomenon of the quantum thermometer without control,
a new peak arises to the left of the original one, highlighting the
precision enhancement and efficient operation-range extension of the
thermometer in the low-temperature region.

Our findings provide an alternative to the precision improvement of
the quantum estimation schemes, irrespective of the origin of the
unknown parameter (from or independent of the bath) and the dynamical
nature of the bath (Markovian or non-Markovian). Moreover, stemming
from the two features of the DM, our proposal may find extensive applications
in estimation processes under noisy channels \citep{2013zhongwei,2013tanqingshou}
as well as the bath sensing tasks \citep{2017bath_sensing,2018bath_sensing,2019bath_sensing,2020bath_sensing,2021bath_sensing}
beyond the two examples we exhibit here.
\begin{acknowledgments}
G.D is supported by National Natural Science Foundation of China (NSFC)
Grant No. 12205211. Y.Y. is supported by NSFC Grant No. 12175204.
\end{acknowledgments}

\appendix

\section{\label{sec:integro-differential equation}The solution of the integro-differential
equation}

Here, we show the derivation of the Eq. (\ref{eq:cet_analy_approx_solution})
in the main text. Inserting the Lorentzian spectral density into the
dynamical equation Eq. (\ref{eq:integro-differential_equation}),
we obtain
\begin{equation}
\dot{c}_{e}(t)=-\Omega^{2}\int_{0}^{t}d\tau c_{e}(\tau)e^{(i\delta_{c}-\frac{\lambda}{2})(t-\tau)}e^{i\xi[\sin(\nu t)-\sin(\nu\tau)]},\label{eq:integro-differential_equation_2-1}
\end{equation}
where we have defined the detuning $\delta_{c}\equiv\omega_{0}-\omega_{c}$.
With the Jacobi-Anger identity $\exp(i\xi\sin(\nu t))=\sum_{n=-\infty}^{\infty}J_{n}(\xi)\exp(in\nu t)$
where $J_{n}(\xi)$ is the $n$th first kind Bessel function, Eq.
(\ref{eq:integro-differential_equation}) can be re-written as
\begin{align}
\dot{c}_{e}(t) & =-\Omega^{2}\sum_{n,m=-\infty}^{\infty}J_{n}(\xi)e^{i(n-m)\nu t}\nonumber \\
 & \times\int_{0}^{t}d\tau c_{e}(\tau)e^{(i\delta_{m}-\frac{\lambda}{2})(t-\tau)},\label{eq:integro-differential_equation_3-1}
\end{align}
where we have used the notation $\delta_{m}\equiv\delta_{c}+m\nu$.
For a very large frequency, i.e., $\nu\gg\Omega,\lambda$, the off-diagonal
terms in Eq. (\ref{eq:integro-differential_equation_3-1}) oscillate
rapidly with time and thus vanish in the long-time limit. Therefore,
Eq. (\ref{eq:integro-differential_equation_3-1}) can be approximated
as
\begin{equation}
\dot{c}_{e}(t)\simeq-\Omega^{2}\sum_{n=-\infty}^{\infty}J_{n}(\xi)\int_{0}^{t}d\tau c_{e}(\tau)e^{(i\delta_{n}-\frac{\lambda}{2})(t-\tau)}.
\end{equation}
After applying the Laplace transformation, we find
\begin{equation}
c_{e}(s)\simeq\frac{c_{e}(0)}{s+\Omega^{2}J_{n_{0}}^{2}(\xi)\frac{1}{s+\frac{\lambda}{2}-i\delta_{n_{0}}}},
\end{equation}
where we only keeps the $n_{0}$th term in the summation which satisfies
$\delta_{n_{0}}\equiv\min\{\left|\delta_{n}\right|,n\in Z\}$.

Finally, we obtain Eq. (\ref{eq:cet_analy_approx_solution}) in the
main text,
\begin{align}
c_{e}(t) & \simeq c_{e}(0)e^{-(\frac{\lambda}{2}-i\delta_{n_{0}})\frac{t}{2}}\nonumber \\
 & \times\left(\cosh\frac{\varTheta_{n_{0}}t}{2}+\frac{\frac{\lambda}{2}-i\delta_{n_{0}}}{\varTheta_{n_{0}}}\sinh\frac{\varTheta_{n_{0}}t}{2}\right),
\end{align}
where we have defined $\varTheta_{n_{0}}\equiv\sqrt{(\lambda/2-i\delta_{n_{0}})^{2}-4\Omega^{2}J_{n_{0}}^{2}(\xi)}$.

\section{\label{sec:Ramsey spectroscopy}the derivative $\partial R(T_{f})/\partial\omega_{0}$
in the long-time region}

\begin{figure}[t]
\begin{centering}
\includegraphics[scale=0.35]{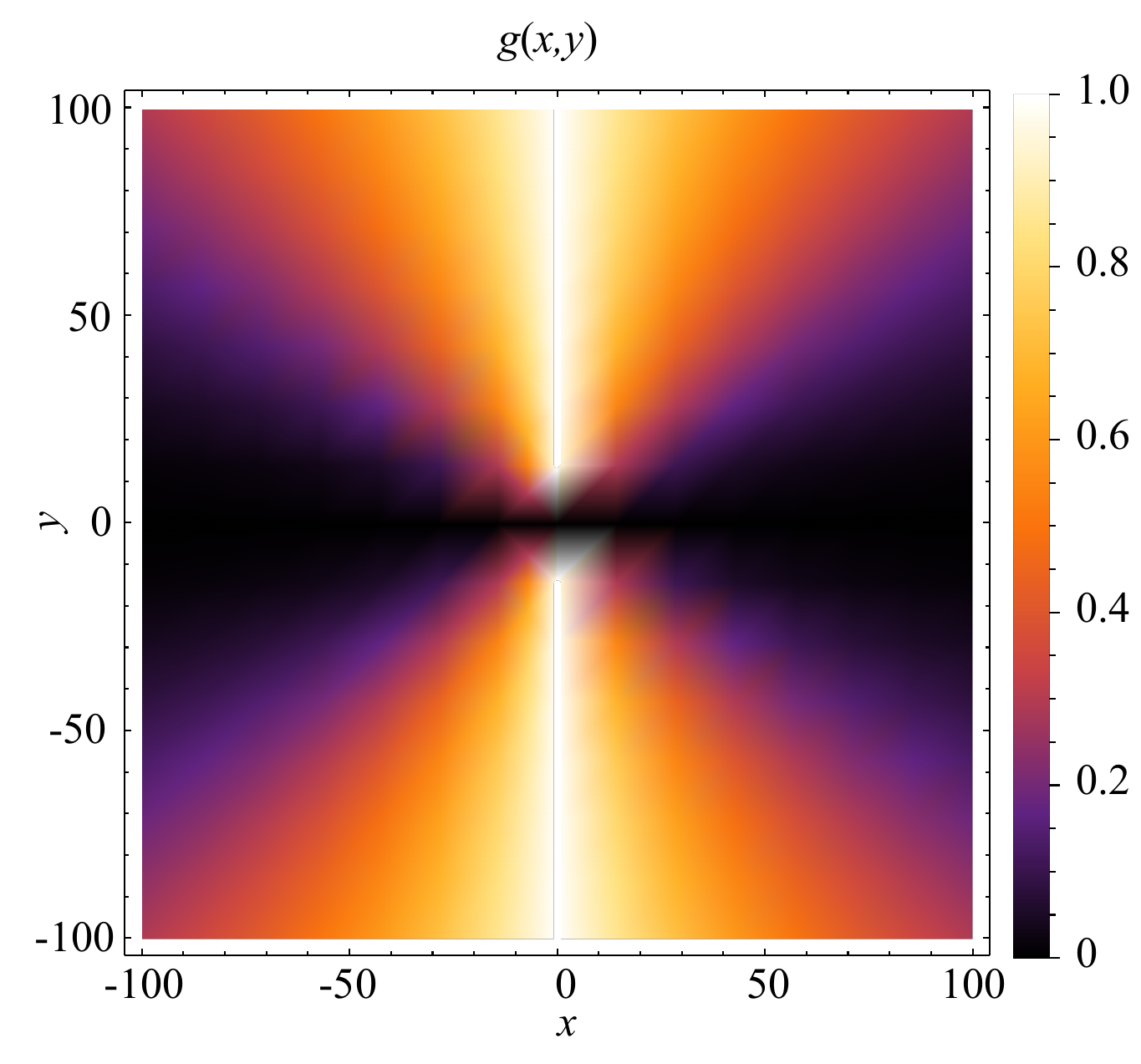}
\par\end{centering}
\caption{\label{fig:gxy}The variation of the function $g(x,y)\equiv1-\left|\mathrm{Re}(\sqrt{(1-ix)^{2}-y^{2}})\right|$
versus $x$ and $y$.}
\end{figure}

\begin{figure}[t]
\begin{centering}
\includegraphics[scale=0.35]{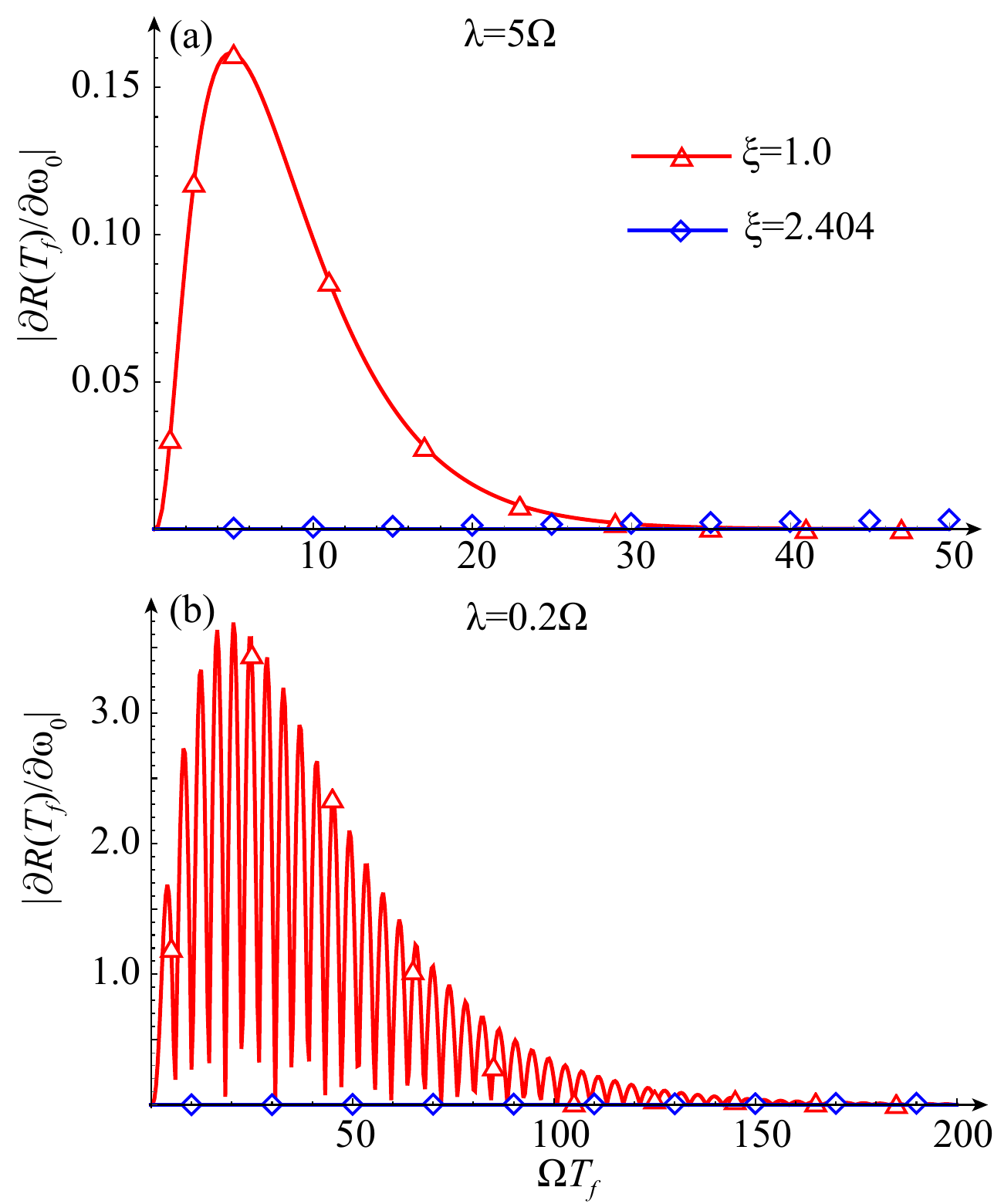}
\par\end{centering}
\caption{\label{fig:derivative Rt}The modulus of the derivative $\partial R(T_{f})/\partial\omega_{0}$
versus time. (a) Markovian: $\lambda=5\Omega$. (b) non-Markovian:
$\lambda=0.2\Omega$. Here the initial state is chosen as $\left|\varphi_{s}(0)\right\rangle =(\left|e\right\rangle +\left|g\right\rangle )/\sqrt{2}$.
The red lines and triangles represent the analytical approximate and
numerical results for the case of $\xi=1$ while the blue lines and
diamonds show that for the case of $\xi=2.404$. Here $\nu=100\Omega$
and $\delta_{c}=0$.}
\end{figure}

Here, we will study the evolution of the derivative $\partial R(T_{f})/\partial\omega_{0}$
and show its vanishing in the long-time region ($T_{f}\gg1$). With
the analytical approximate expression Eq. (\ref{eq:cet_analy_approx_solution}),
the derivative of $R(T_{f})$ over $\omega_{0}$ becomes ($\partial R(T_{f})/\partial\omega_{0}=c_{e}(0)^{-1}\partial c_{e}(T_{f})/\partial\omega_{0}$)

\begin{align}
\frac{\partial R(T_{f})}{\partial\omega_{0}} & \simeq ie^{-\left(\frac{\lambda}{4}-i\frac{\delta_{n_{0}}}{2}\right)T_{f}}\frac{4\Omega^{2}J_{n_{0}}^{2}(\xi)}{\varTheta_{n_{0}}^{3}}\sinh\frac{\varTheta_{n_{0}}T_{f}}{2}\nonumber \\
 & +ie^{-\left(\frac{\lambda}{4}-i\frac{\delta_{n_{0}}}{2}\right)T_{f}}\frac{T_{f}}{2}\nonumber \\
 & \times\left(1-\frac{(\frac{\lambda}{2}-i\delta_{n_{0}})^{2}}{\varTheta_{n_{0}}^{2}}\right)\cosh\frac{\varTheta_{n_{0}}T_{f}}{2}\nonumber \\
 & =i\frac{4\Omega^{2}J_{n_{0}}^{2}(\xi)}{\varTheta_{n_{0}}^{3}}e^{-\left(\frac{\lambda}{4}-i\frac{\delta_{n_{0}}}{2}\right)T_{f}}\nonumber \\
 & \times\left(\sinh\frac{\varTheta_{n_{0}}T_{f}}{2}-\frac{\varTheta_{n_{0}}T_{f}}{2}\cosh\frac{\varTheta_{n_{0}}T_{f}}{2}\right).\label{eq:derivative Rt appr}
\end{align}
We find that when the modulation amplitude $\xi$ is chosen as one
of the zeros of the Bessel function $J_{n_{0}}(\xi)$, Eq. (\ref{eq:derivative Rt appr})
vanishes. In the other cases where $J_{n_{0}}(\xi)\neq0$, when the
free-evolution time $T_{f}$ is large, the second term in the brackets
of Eq. (\ref{eq:derivative Rt appr}) dominates. Hence, Eq. (\ref{eq:derivative Rt appr})
can be further simplified as
\begin{align}
 & \frac{\partial R(T_{f})}{\partial\omega_{0}}\nonumber \\
 & \simeq-i\frac{\Omega^{2}J_{n_{0}}^{2}(\xi)T_{f}}{\varTheta_{n_{0}}^{2}}e^{-\left(\frac{\lambda}{4}-i\frac{\delta_{n_{0}}}{2}\right)T_{f}}\left(e^{\frac{\varTheta_{n_{0}}T_{f}}{2}}+e^{-\frac{\varTheta_{n_{0}}T_{f}}{2}}\right).\label{eq:derivative Rt appr-1}
\end{align}
Now the behavior of the derivative in the long-time limit depends
on the real part of $\varTheta_{n_{0}}$. For the case of $\left|\mathrm{Re}(\varTheta_{n_{0}})\right|<\lambda/2$,
the derivative is zero. When $\left|\mathrm{Re}(\varTheta_{n_{0}})\right|>\lambda/2$,
the derivative diverges. To figure out the relations between $\lambda/2$
and $\left|\mathrm{Re}(\varTheta_{n_{0}})\right|$, we define a function
$g(x,y)\equiv1-\left|\mathrm{Re}(\sqrt{(1-ix)^{2}-y^{2}})\right|$
and plot it numerically in Fig. \ref{fig:gxy} for a wide range of
$x$ and $y$. It is demonstrated that $g(x,y)$ remains nonnegative
in all region, indicating the vanishing of the derivative $\partial R(T_{f})/\partial\omega_{0}$
in the long-time limit.

Figure \ref{fig:derivative Rt} illustrates the dynamics of the modulus
of the derivative ($\left|\partial R(T_{f})/\partial\omega_{0}\right|$)
in both the Markovian and non-Markovian regimes. We find that our
approximate analysis (Eq. (\ref{eq:derivative Rt appr})) is in good
agreement with the numerical results (Eq. (\ref{eq:integro-differential_equation})
in the main text). In the case where $\xi$ is one of the zeros of
$J_{n_{0}}(\xi)$, $\left|\partial R(T_{f})/\partial\omega_{0}\right|$
remains vanishing for all $T_{f}$ while for $J_{n_{0}}(\xi)\neq0$
the envelope of the derivative $\left|\partial R(T_{f})/\partial\omega_{0}\right|$
first increases with time and then decreases to zero in the long-time
limit.

\section{The second-order TCL master equation \label{sec:appen_TCL_equation}}

Here we derive the second-order TCL master equation. Up to the second
order, the master equation becomes
\begin{equation}
\dot{\rho}_{s}(t)\simeq-\int_{0}^{t}Tr_{b}\left([H_{I}(t),[H_{I}(\tau),\rho_{s}(\tau)\otimes\rho_{b}]]\right)d\tau,\label{eq:2 order master equ}
\end{equation}
where the Born approximation has been used. In the TCL method, we
ignore the convolution of the reduced density matrix $\rho_{s}(t)$
on the right of Eq. (\ref{eq:2 order master equ}) by replacing $\tau$
with $t$. Inserting the Hamiltonian in the interaction picture (Eq.
(\ref{eq:Interaction Hamil})), we obtain

\begin{widetext}

\begin{align}
\dot{\rho}_{s}(t) & \simeq-\int_{0}^{t}Tr_{b}\left([H_{I}(t),[H_{I}(\tau),\rho_{s}(t)\otimes\rho_{b}]]\right)d\tau\nonumber \\
 & =\sum_{k,k^{'}}\int_{0}^{t}Tr_{b}\left\{ \left[g_{k}\sigma_{+}b_{k}e^{i(\omega_{0}-\omega_{k})t+i\xi\sin(\nu t)}+h.c.\right]\rho_{s}(t)\otimes\rho_{b}\left[g_{k^{'}}^{*}\sigma_{-}b_{k^{'}}^{\dagger}e^{-i(\omega_{0}-\omega_{k^{'}})\tau-i\xi\sin(\nu\tau)}+h.c.\right]+h.c.\right\} d\tau\nonumber \\
 & -\sum_{k,k^{'}}\int_{0}^{t}Tr_{b}\left\{ \left[g_{k}\sigma_{+}b_{k}e^{i(\omega_{0}-\omega_{k})t+i\xi\sin(\nu t)}+h.c.\right]\left[g_{k^{'}}^{*}\sigma_{-}b_{k^{'}}^{\dagger}e^{-i(\omega_{0}-\omega_{k^{'}})\tau-i\xi\sin(\nu\tau)}+h.c.\right]\rho_{s}(t)\otimes\rho_{b}+h.c.\right\} d\tau\nonumber \\
 & =\int_{0}^{\infty}d\omega J(\omega)\int_{0}^{t}d\tau N_{\pm}(\omega)\left\{ \left[e^{i\phi(t)-i\phi(\tau)}+c.c.\right]\sigma_{+}\rho_{s}(t)\sigma_{-}-\sigma_{-}\sigma_{+}\rho_{s}(t)e^{-i\phi(t)+i\phi(\tau)}-\rho_{s}(t)\sigma_{-}\sigma_{+}e^{i\phi(t)-i\phi(\tau)}\right\} \nonumber \\
 & +\int_{0}^{\infty}d\omega J(\omega)\int_{0}^{t}d\tau\left[1\mp N_{\pm}(\omega)\right]\left\{ \left[e^{-i\phi(t)+i\phi(\tau)}+c.c.\right]\sigma_{-}\rho_{s}(t)\sigma_{+}-\sigma_{+}\sigma_{-}\rho_{s}(t)e^{i\phi(t)-i\phi(\tau)}-\rho_{s}(t)\sigma_{+}\sigma_{-}e^{-i\phi(t)+i\phi(\tau)}\right\} .
\end{align}

\end{widetext}where we have defined $\phi(t)\equiv(\omega_{0}-\omega)t+\xi\sin\nu t$.
$N_{\pm}(\omega)\equiv(e^{\omega/T}\pm1)^{-1}$ stands for the mean
particle number for the fermionic (plus sign) and bosonic (minus sign)
bath. Hence, for the fermionic bath, the diagonal and off-diagonal
elements of $\rho_{s}(t)$ becomes

\begin{align}
\dot{\mathrm{P}}_{e}(t) & \simeq\int_{0}^{t}d\tau\int_{0}^{\infty}d\omega J(\omega)2\cos\left[\phi(t)-\phi(\tau)\right]\nonumber \\
 & \times\left[N_{+}(\omega)-\mathrm{P}_{e}(t)\right],\nonumber \\
\dot{\rho}_{eg}(t) & \simeq-\int_{0}^{t}d\tau\int_{0}^{\infty}d\omega J(\omega)e^{i\phi(t)-i\phi(\tau)}\rho_{eg}(t).\label{eq:TCl_equa-1}
\end{align}
which is Eq. (\ref{eq:TCl_equa}) in the main text.

\section{The bosonic bath case\label{sec:appen-bosonic-bath}}

\begin{figure*}[t]
\begin{centering}
\includegraphics[scale=0.3]{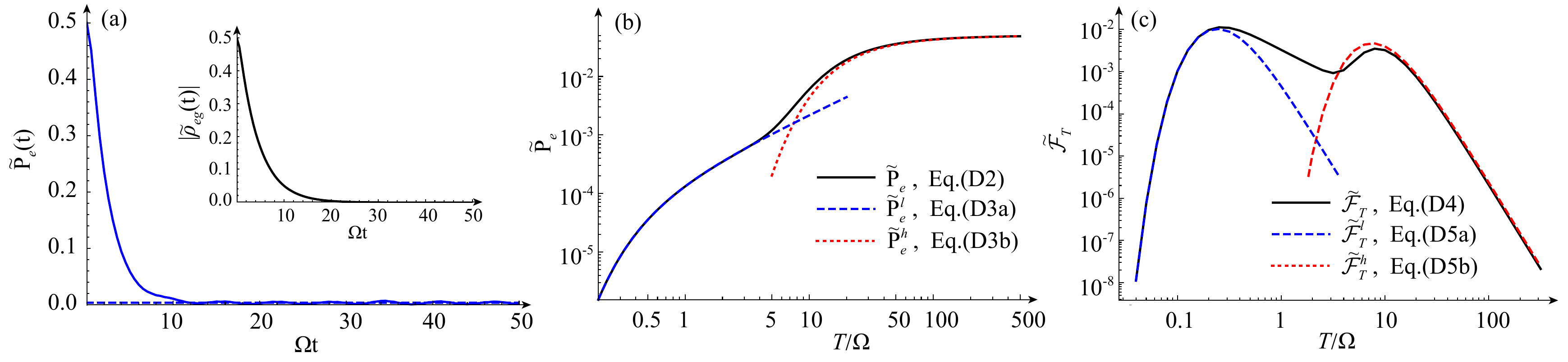}
\par\end{centering}
\caption{\label{fig:boson bath}(a) The dynamics of the probability in the
excited state $\mathrm{\widetilde{\mathrm{P}}}_{e}(t)$ under the
Markovian dynamics ($\lambda=5\Omega$) with the temperature $T=2\Omega$.
The inset gives the dynamics of the off-diagonal term $\left|\widetilde{\rho}_{eg}(t)\right|$.
The initial state of the TLS is chosen as $\left|\varphi_{s}(0)\right\rangle =(\left|e\right\rangle +\left|g\right\rangle )/\sqrt{2}$.
(b) The steady-state probability in the excited state $\mathrm{\widetilde{\mathrm{P}}}_{e}$
and (c) the Fisher information $\mathcal{\widetilde{F}}_{T}$ versus
temperature. The black lines plot the analytical results while the
blue dashed (red dotted) lines show the analytical approximation in
the low-temperature (high-temperature) region. Here $\omega_{0}=31\Omega$,
$\delta_{c}=0$, $\nu=30\Omega$, and $\xi=1.0$.}
\end{figure*}

In this section, we demonstrate the robustness of our proposal in
the bosonic bath.The dynamical equations of the reduced density matrix
elements read
\begin{align}
\mathrm{\dot{\widetilde{P}}}_{e}(t) & =\int_{0}^{\infty}d\omega J(\omega)\int_{0}^{t}d\tau\left\{ e^{i\left[\phi(t)-\phi(\tau)\right]}+c.c.\right\} \nonumber \\
 & \times\left\{ N_{-}(\omega)-\left[2N_{-}(\omega)+1\right]\widetilde{\mathrm{P}}_{e}(t)\right\} ,\nonumber \\
\dot{\widetilde{\rho}}_{eg}(t) & =-\int_{0}^{\infty}d\omega J(\omega)\int_{0}^{t}d\tau\left[2N_{-}(\omega)+1\right]\nonumber \\
 & \times\widetilde{\rho}_{eg}(t)e^{i\left[\phi(t)-\phi(\tau)\right]},
\end{align}
where $\mathrm{\widetilde{\mathrm{P}}}_{e}(t)$ ($\widetilde{\rho}_{eg}(t)$)
represents the diagonal (non-diagonal) term of the reduced density
matrix in the case of the bosonic bath. Similar to the fermionic case,
in the long-time limit, the off-diagonal term $\widetilde{\rho}_{eg}(t)$
vanishes and the diagonal term $\mathrm{\widetilde{\mathrm{P}}}_{e}(t)$
reaches a steady-state value,
\begin{align}
\mathrm{\widetilde{\mathrm{P}}}_{e} & =\sum_{n}\widetilde{P}_{n}N_{-}(\omega_{n})\nonumber \\
 & =\frac{\sum_{n}J_{n}^{2}(\xi)J(\omega_{n})N_{-}(\omega_{n})}{\sum_{n}J_{n}^{2}(\xi)J(\omega_{n})(1+2N_{-}(\omega_{n}))}\label{eq:Pe_ss_Boson}
\end{align}
Therefore, the steady-state probabilities in the low-temperature and
high-temperature regions read

\begin{subequations}
\begin{align}
\mathrm{\widetilde{\mathrm{P}}}_{e}^{l} & \simeq\widetilde{P}_{n_{1}}N_{-}(\omega_{n_{1}}),\\
\widetilde{\mathrm{P}}_{e}^{h} & \simeq N_{+}(\omega_{n_{0}}).\label{eq:Pe_l=000026h_Boson}
\end{align}

\end{subequations} Using Eq. (\ref{eq:Pe_ss_Boson}), we obtain the
QFI

\begin{equation}
\mathcal{\widetilde{F}}_{T}=\frac{1}{T^{4}}\frac{\left(\frac{1}{4}\sum_{n}\widetilde{P}_{n}\omega_{n}\frac{1}{\sinh^{2}(\omega_{n}/2T)}\right)^{2}}{\mathrm{\widetilde{\mathrm{P}}}_{e}\left(1+\mathrm{\widetilde{\mathrm{P}}}_{e}\right)\left(1+2\mathrm{\widetilde{\mathrm{P}}}_{e}\right)^{2}}.\label{eq:FI_Boson}
\end{equation}

In the low-temperature and high-temperature regions, the QFI becomes

\begin{subequations}
\begin{align}
\mathcal{\mathcal{\widetilde{F}}}_{T}^{l} & \simeq P_{n_{1}}\frac{\omega_{n_{1}}^{2}}{4T^{4}}\frac{1}{\sinh^{2}(\omega_{n_{1}}/2T)},\\
\mathcal{\mathcal{\widetilde{F}}}_{T}^{h} & \simeq F_{T}^{0}(\omega_{n_{0}}).\label{eq:FI_l=000026h_Boson}
\end{align}
\end{subequations}

Figure \ref{fig:boson bath} displays the result of the bosonic bath.
Analogous to the fermionic bath case, in the steady state, there are
two effective frequencies of the thermometer corresponding to the
low- and high-temperature regions, respectively. Moreover, for a careful
choice of the modulation parameter, a new peak of the QFI emerges
to the left of the original one, indicating the broadening of the
efficient operation region of the quantum thermometer.

\end{document}